\documentclass{IEEEtran}

\usepackage{graphicx}
\usepackage{amsmath}
\usepackage{amssymb}
\usepackage{cite}
\usepackage{tabularx}
\usepackage{caption}
\usepackage{subcaption}
\usepackage{balance}
\usepackage{color}
\usepackage{mathtools}
\usepackage{url}
\usepackage{multirow}
\usepackage{placeins}
\usepackage{subcaption}
\usepackage{color}
\usepackage{enumitem}
\IEEEoverridecommandlockouts

\newcommand{\newlinecell}[2][c]{%
      \begin{tabular}[#1]{@{}c@{}}#2\end{tabular}}

\setlist[description]{font=\normalfont}

\title{Massive MIMO Optimization with\\Compatible Sets}

\author{\IEEEauthorblockN{Emma Fitzgerald\IEEEauthorrefmark{1}
\quad\and
Micha{\l} Pi\'{o}ro\IEEEauthorrefmark{2}
\quad\and
Fredrik Tufvesson\IEEEauthorrefmark{3}}\\[0.5em]
\IEEEauthorrefmark{1}\IEEEauthorrefmark{3}Department of Electrical and
Information Technology, Lund University,\\SE-221 00 Lund, Sweden\\[0.5em]
\IEEEauthorrefmark{1}\IEEEauthorrefmark{2}Institute of Telecommunications, Warsaw University of
Technology,\\Nowowiejska 15/19, 00-665 Warsaw, Poland\\[0.5em]
\IEEEauthorblockA{
\IEEEauthorrefmark{1}emma.fitzgerald@eit.lth.se \IEEEauthorrefmark{2}m.pioro@tele.pw.edu.pl
\IEEEauthorrefmark{3}fredrik.tufvesson@eit.lth.se}
}

\begin{document}

\maketitle

\begin{abstract}
Massive multiple-input multiple-output (MIMO) is expected to be a vital
component in future 5G systems. As such, there is a need for new modeling
in order to investigate the performance of massive MIMO not only at the
physical layer, but also higher up the networking stack. In this paper, we
present general optimization models for massive MIMO, based on mixed-integer
programming and compatible sets, with both maximum ratio combing and zero
forcing precoding schemes. We then apply our models to the case of joint
device scheduling and power control for heterogeneous devices and traffic
demands, in contrast to existing power control schemes that consider only
homogeneous users and saturated scenarios. Our results show substantial
benefits in terms of energy usage can be achieved without sacrificing
throughput, and that both signalling overhead and the complexity of end
devices can be reduced by abrogating the need for uplink power control
through efficient scheduling.
\end{abstract}

\allowdisplaybreaks

\section{Introduction}

Massive multiple-input multiple-output (MIMO) is slated to be a key enabling
technology for 5G \cite{gupta2015survey,boccardi2014five}, in particular in
order to meet the ambitious goal of a thousandfold increase in data traffic
\cite{chen2014requirements}. For the enhanced mobile broadband use case, where
the required increase in capacity is induced by relatively few users executing
demanding applications, the ability of massive MIMO systems to serve many users
simultaneously with the same frequency resources is a clear advantage. However,
in the massive machine-type communication case, it becomes impossible even for a
massive MIMO base station to serve all devices at once, leading to new resource
allocation problems in how to assign the limited number of available pilot
signals to devices \cite{de2017random}. This is particularly challenging in cases
where the devices are heterogeneous, with differing traffic volumes and quality of
service requirements.

For massive machine-type communications, arising for example in Internet of
Things (IoT) use cases, where the number of users per cell is larger than the
number of pilot signals, it is not possible for each device to be assigned a
pilot. If traffic demands are sporadic, then random access protocols may be the
the most appropriate access method, and some work has been conducted on
designing such protocols for massive MIMO capable of resolving pilot collisions
\cite{bjornson2016random,sorensen2014massive,de2016random}. However, these
protocols suffer from the usual inefficiencies present in any random access
protocol; in particular the above protocols require that devices transmit multiple
copies of data in order to resolve collisions through successive interference
cancellation, resulting in less efficient use of resources. For high offered
load scenarios and/or those with less variable traffic demands --- especially in the most
extreme case of periodic traffic --- scheduled access may be more appropriate.

Regardless of the access protocol used, there is a need for new models based on
the physical layer technology of massive MIMO, but appropriate for investigating
higher-layer performance. In this paper we develop such a model, based on
mixed-integer programming, and apply it to the case of end device scheduling for
maximal throughput, with heterogeneous traffic demands (different data volumes
to transmit and different data rates for different users). Here, we optimize not
only the allocation of pilot resources to devices, but also transmission power
control on both the uplink and downlink, and investigate whether it is possible
to avoid the need for complex power control schemes through efficient scheduling
of end device transmissions. We consider two widely-used precoding schemes:
maximum ratio combining (MRC) and zero forcing (ZF). These can be seen as the
two corner cases for precoding: one where no regard is paid to minimizing mutual
interference but rather only maximizing receiver power levels for all uses, and
one where mutual interference is completely nulled out (in the ideal case),
respectively. In practice, a compromise such as minimum mean-square error or
regularized zero-forcing is used, but it is still of interest to investigate the
two extreme cases. We also consider three different methods for power control.
We have performed systematic numerical studies to investigate the performance of
our models, in terms of solution time, achievable throughput, and energy usage.

The contributions of this paper are the following:
\begin{itemize}
    \item We present new optimization models suitable for massive MIMO systems
    	that allow for heterogeneous users and traffic, as well as a greater
    	number of users than available pilots, as is typical of IoT scenarios.
    	In particular, we provide formulations based on compatible sets, previously
    	used for optimization of wireless mesh networks, based on the
    	massive MIMO physical layer.
    \item We develop an efficient solution approach based on column generation,
    	capable of handling complex problems with non-compact formulations
    	and integer decision variables. Both our models and solution approach
    	are general and can be applied to a wide range of different objectives.
    \item We provide formulations for two widely used precoding schemes
	(maximum ratio combining and zero forcing), as well as three power
	control schemes. These are the existing max-min fair SINR power control
	scheme, a new version of this scheme taking into account scheduling, and
	fully optimised power control. Further, we investigate a variant of fair
	power control performed only on the downlink.
    \item We apply our models and solution approach to the problem of joint
    	scheduling and power control for throughput maximization for a large
    	number of heterogeneous devices (more than the available pilots). We
    	conduct an extensive numerical study to investigate and compare the
    	performance of the different precoding and power control schemes, as
    	well as of the optimization itself. In doing so, we identify which cases
    	are most suitable for different scheduling and power control methods.
    \item We create an efficient heuristic that provides similar performance to
    	full optimization, while substantially reducing the solution time of the
    	main integer programming problem.
    \item We show that for IoT scenarios, where each device does not aim for
    	maximum throughput but rather has its own traffic and QoS parameters,
    	the existing power control methods are much worse for energy efficiency
    	than when the transmission power is optimized taking into account each
    	device's traffic demand and desired rate. Further, we show that joint
    	optimization of power control and scheduling greatly improves throughput
    	in cases where the channel quality of devices is unbalanced, with a
    	small group of devices with good channels, and a larger group with poor
    	channels.
    \item We show that it is possible to achieve the same performance in terms
    	of throughput without power control on the uplink by scheduling devices
    	efficiently. This saves significant signalling overhead and complexity
    	on end devices, which is important for resource-constrained IoT devices.
\end{itemize}
The rest of this paper is organized as follows. Section \ref{sec:related_work}
describes related work in this area. Section \ref{sec:models} elaborates our
targeted scenario and system models. Section \ref{sec:problem-solution} then
details our optimization problems and solution approach. A discussion of the
complexity and efficiency of our approach follows in Section
\ref{sec:efficiency}. In Section \ref{sec:experiments} we present our numerical
study, along with results and discussion. In Section \ref{sec:future_work} we
discuss future work, and, finally, in Section \ref{sec:conclusion} we conclude
this paper.

\section{Related Work}\label{sec:related_work}

Massive MIMO, first proposed in \cite{marzetta2010noncooperative}, refers to
multiple-antenna deployments in which the number of antennas at the base station
is significantly higher than the number of user antennas. This allows us to
exploit favorable propagation, that is, that the channel responses from each
antenna at the base station to the different user terminals are sufficiently
different to allow separation of the users' data streams by digital pre- and
post-processing. In the presence of favorable propagation and a large number of
antennas, an effect known as channel hardening \cite{larsson2014massive} arises,
and a radical increase in spectral efficiency is possible
\cite{harris2016serving}. Channel hardening means that each channel will be
close to its expected value and channel variation is negligible in both the time
and frequency domains. The increase in capacity gained from using massive MIMO
thus comes as a direct result of considerably increasing the number of antennas
and benefiting from the statistical advantages derived from the resulting large
number of different channels to each user.

Channel models for massive MIMO can be divided into two categories:
correlation-based models and geometry-based models \cite{zheng2014massive,
araujo2016massive}. Geometry-based models
\cite{payami2012channel,gao2012measured} can be used for the performance
evaluation of practical systems, while theoretical performance analysis of these
systems often relies on correlation-based models. The latter type includes
correlation channel models \cite{hoydis2013massive,yin2013coordinated}, where
the correlation between antennas is considered, and mutual coupling channel
models \cite{masouros2013large}, where the coupling between each pair of
antennas is also taken into account. Perhaps the mostly widely used models for
performance analysis are however independent and identically distributed
Rayleigh fading models
\cite{marzetta2010noncooperative,rusek2013scaling,hochwald2004multiple}. Here,
small-scale fading of the channel is modeled by i.i.d. Gaussian variables. This
makes the analysis more tractable but nonetheless yields models that are sufficiently
powerful --- despite their lack of realism --- to demonstrate important results
for massive MIMO systems, including the aforementioned channel hardening property
\cite{larsson2014massive}, as well as the effective SINR expressions we will use
in this paper.

While the above models include small-scale fading of the channel in individual
coherence blocks, the effective SINR characterizes the expected performance over
a larger number of blocks, and has been the subject of research since the early
days of massive MIMO
\cite{hoydis2011massive,hoydis2013massive,ngo2013energy,yang2013performance,bjornson2016massiveb}.
In our work, we are primarily interested in performance of massive MIMO systems
over a larger time scale, and as such we take the effective SINR as the basis
for our models. In particular, the effective SINR for the uplink in single cell
systems was derived in \cite{ngo2013energy}, while for the downlink it was
analyzed in \cite{yang2013performance}. Although effective SINR results are also
available for multi-cell massive MIMO systems, we do not consider these in this
paper.

In our models, we also take into account transmission power control for both the
uplink and downlink. Most existing work on power control for massive MIMO
systems considers homogeneous users. Max-min fair power control
\cite{bjornson2016massivea}, in which the minimum SINR amongst the users is
maximized, is a commonly adopted scheme. Although not true of max-min
optimization in general, in the case of massive MIMO power control, this results
in a common SINR value for all users. Maximum (fair) SINR is however not the
only objective considered in previous work. Power control for mitigation of pilot
contamination, in which reuse of pilot signals (usually in different cells)
results in impaired channel estimation, has been studied in
\cite{saxena2015mitigating}, and energy efficiency has been considered in
\cite{zappone2016energy}. In \cite{choi2014massive}, joint power control is
performed between cells, and here users are not homogeneous, but rather each
user has an individual SINR target. However, different traffic demands for
users are not considered. In our work here, we combine power control with user
scheduling, in which each user may have a defined individual SINR target and
demand, that is, the amount of data the user wishes to send.

We achieve the above by applying optimization, specifically mixed-integer
programming. Optimization methods have already seen use for various purposes
within the area of massive MIMO systems. In \cite{bogale2014pilot}, weighted-sum
mean-square error minimization and Rayleigh quotient methods were used to
determine pilot signals that reduce pilot contamination and improve channel
estimation. For the case where the number of pilots is equal to the number of
users, semi-definite programming and convex optimization were used to find
optimal solutions. User scheduling in the case where the number of users exceeds
the number of pilots was not considered, however. In
\cite{bjornson2015optimal}, the optimal number of antennas, number of active
users, and transmission power were optimized for maximal energy efficiency. The
focus was on studying the performance of massive MIMO systems, and
closed-form expressions were obtained for the above. However, in order to do so,
each parameter was considered one at a time, while the other two were held
constant. We instead aim to develop methods for user scheduling and power
control for specific scenarios, and consider joint optimization of the two,
since the transmission power affects which users are able to transmit or receive
simultaneously.

Joint power optimization and user association in multi-cell systems was studied
in \cite{van2016joint}, where each user was associated to a set of base
stations that would then serve that user. The models provided consist of
efficient linear programming formulations, however all users are always able to
associate to at least one base station, that is, there is no limit placed on the
number of available pilot signals. This means that user groups are static ---
only a single association was performed, rather than dynamic scheduling ---
resulting in a smaller solution space. and since user selection is not required,
no integer variables are needed. For the scenario we consider here, with many more
users than pilot signals, mixed-integer programming formulations are needed, and
we use column generation to deal with the large number of possible combinations
of simultaneously transmitting users.

Fair scheduling in multi-cell systems was investigated in \cite{huh2011multi}
with asymptotic analysis using large random matrix theory and convex
optimization. However, no solution was provided for actually
scheduling the users, but rather the achievable fair rate was analyzed. Further, the
analysis relies on assumptions that we do not require in our work, namely that
the ratio of antennas to users is kept constant, and that users are scheduled in
co-located groups. Joint antenna and user selection has also been studied for
massive MIMO, however using only brute-force search \cite{guozhen2014joint},
which has very high computation complexity and is thus impractical for all but
very small problem instances, and a greedy algorithm \cite{benmimoune2015joint},
which is more efficient but provides only suboptimal solutions in general.
Finally, user grouping and scheduling has also been studied in
frequency-division duplex (FDD) massive MIMO systems \cite{xu2014user}, however
users were placed in pre-beamforming groups, and only users in the same
group could be scheduled together. These groups were formed using clustering
algorithms that do not provide optimal solutions, and user scheduling was
performed based on an SINR approximation that considers only a single user at a
time. Moreover, time-division duplex (TDD) provides better performance for
massive MIMO than FDD \cite{bjornson2016massivea}, and as such is a more
suitable candidate for real implementations. In our work we consider a
TDD-based system.

Our optimization models presented in this paper are based on the notion of
compatible sets (c-sets), which were first introduced in \cite{Bjorklund2003}
for transmission scheduling in wireless mesh networks. Wireless mesh networks,
like massive MIMO systems but unlike previous generations of cellular systems,
allow multiple, simultaneous, possibly interfering transmissions. A compatible
set is then a set of simultaneous transmissions that are able to be successfully
decoded at the receivers despite this possible interference, that is, where the
SINR is sufficiently high at all receivers.

Since their introduction, compatible sets have seen many applications in
optimization of wireless mesh networks. Extensions for multiple modulation and coding
schemes (MCSs) and power control were given in \cite{CaC2006} and
\cite{Pioro2011}. C-sets have been applied to joint link rate assignment and
transmission scheduling \cite{Yuan-CS-heu-2015}, link scheduling
\cite{Angelakis2014}, routing and scheduling for throughput optimization
\cite{Capone2010}, and multicast routing and scheduling
\cite{PTC-2018}. While the above were primarily focused on
throughput maximization, work using c-sets has also considered fairness
\cite{Pioro2011,Zotkiewicz2014}, delay minimization
\cite{Capone2014,Capone2015-inoc} and energy efficiency
\cite{fitzgerald2018energy}.

The notion of compatible sets thus provides a general and flexible method for
modeling and solving optimization problems for wireless systems. However,
existing models for c-sets are based on nodes equipped with omnidirectional
antennas where interference depends primarily on the distance between receivers
and transmitters. In massive MIMO systems, interference in simultaneous
transmissions has different causes, for example imperfect channel estimation,
and so these models cannot be applied to massive MIMO in their current form. In
this paper, we develop a new type of compatible set model suitable for massive
MIMO systems. We apply it to the specific case of user scheduling and power
control, however the model we present here is general and can be readily used
for other types of objective functions.

\section{System Models}\label{sec:models}

In the following, we use the channel models and effective SINR expressions
derived in \cite{marzetta2016fundamentals}, Chapter 3, based on the work in
\cite{ngo2013energy} and \cite{yang2013performance}.  The notation used is
summarized in Table \ref{tab:notation}.  We consider a scenario with a
single-cell massive MIMO system. There is one base station with an antenna array
with $M$ antennas, and there is a set $\mathcal K$ of single-antenna end devices
in the cell, with the number of devices $|\mathcal K| = K$. We specifically
consider the case where $K$ is larger than the number of devices that can be
spatially multiplexed by the base station. This will
typically be the case in IoT scenarios, especially for machine-type
communication. The number of devices that can be served simultaneously is
limited by the number of available pilot signals, as well as the required SINR
for the devices being served. Here, we will not address mobility of the end
devices, so each end device $k \in K$ has a fixed location.

\begin{table}
    \begin{centering}
        \begin{tabularx}{\columnwidth}{|c|X|}
            \hline
	    $M$ & number of antennas in the base station's antenna array \\
	    \hline
	    $\mathcal K$ & set of end devices \\
	    \hline
	    $K$ & number of end devices, $K = |\mathcal K|$ \\
	    \hline
	    $r(k)$ & distance of device $k \in \mathcal K$ from the base station \\
	    \hline
	    $T$ & duration of coherence block in seconds \\
	    \hline
	    $B$ & bandwidth of coherence block in Hertz \\
	    \hline
	    $\tau$ & number of samples in each coherence block \\
	    \hline
	    $\tilde{\tau}$ & number of pilot samples in each coherence block \\
	    \hline
	    $P$ & number of available pilots in each coherence block \\
	    \hline
	    $S$ & length of each pilot, in samples\\
	    \hline
	    $\hat{\tau}$ & number of uplink data samples in each coherence block \\
	    \hline
	    $\check{\tau}$ & number of downlink data samples in each coherence block \\
	    \hline
	    $\beta(k)$ & large-scale effects coefficient of end device $k \in
	    \mathcal K$ \\
	    \hline
	    $R$ & reference distance in meters \\
	    \hline
	    $\alpha$ & path loss exponent \\
	    \hline
	    $\gamma(k)$ & mean-square channel estimate for end device $k \in
	    \mathcal K$ \\
	    \hline
	    $\hat{\rho}$ & uplink SNR \\
	    \hline
	    $\check{\rho}$ & downlink SNR \\
	    \hline
	    $\hat{h}(k)$ & device $k$'s uplink demand in coherence blocks, $k
	    \in \mathcal K$ \\
	    \hline
	    $\check{h}(k)$ & device $k$'s downlink demand in coherence blocks, $k
	    \in \mathcal K$ \\
	    \hline
	    $\mu(k)$ & SINR threshold for successful reception for device $k \in
	    \mathcal K$ \\
	    \hline
	    $\mathcal C$ & family of all compatible sets \\
	    \hline
	    $\hat{\mathcal C}(k)$ & family of compatible sets in which end device $k
	    \in \mathcal K$ is a transmitter \\
	    \hline
	    $\check{\mathcal C}(k)$ & family of compatible sets in which end device $k
	    \in \mathcal K$ is a receiver \\
	    \hline
	    $t(c)$ & set of devices $k \in \mathcal K$ that transmit in compatible set $c \in \mathcal C$ \\
	    \hline
	    $r(c)$ & set of devices $k \in \mathcal K$ that receive in compatible set $c \in \mathcal C$ \\
	    \hline
	    $T_c$ & number of coherence blocks assigned to compatible set $c \in \mathcal
	    C$. \\
	    \hline
	    $\hat\pi_k,\,\check\pi_k$ & dual decision variable associated with end device $k \in
	    \mathcal K$ \\
	    \hline
	    $u_k,\,\hat u_k, \, \check u_k$ & whether or not end device $k \in K$ is active in a generated
	    compatible set. \\
	    \hline
	    $\hat{\eta}_k$ & uplink power control coefficient for end device $k \in
	    \mathcal K$ \\
	    \hline
	    $\check{\eta}_k$ & downlink power control coefficient for end device $k \in
	    \mathcal K$ \\
	    \hline
	    $\mathbb{B}$ & the set $\{0, 1\}$ \\
	    \hline
	    $\mathbb{Z}, \mathbb{Z_+}$ & the set of all, and all non-negative,
	    integers, respectively \\
	    \hline
	    $\mathbb{R}, \mathbb{R_+}$ & the set of all, and all non-negative,
	    real numbers, respectively \\
	    \hline
        \end{tabularx}
        \caption{Summary of notation.}
        \label{tab:notation}
    \end{centering}
\end{table}

\subsection{Channel Model}
The channel for the massive MIMO cell can be divided into coherence blocks,
where each coherence block is of duration $T$ s and bandwidth $B$ Hz. This gives
$\tau = T B$ samples, taken at intervals of $1/B$ seconds, in each coherence
block (see \cite{marzetta2016fundamentals}, Section 2.1.3 for further details
--- note that samples as used here are not equivalent to OFDM samples). Of these
samples, $\tilde{\tau}$ are used for pilot transmission, $\hat{\tau}$ for uplink
data transmission, and $\check{\tau}$ for downlink data transmission. We thus
have $\tau = \tilde{\tau} + \hat{\tau} + \check{\tau}$. In each coherence block,
channel state information (CSI) is obtained for each scheduled end device by the
device sending a pilot signal of length $S$ samples. Using the example scheme
for orthogonal pilots given in \cite{marzetta2016fundamentals}, Section 3.1.1,
we thus have a total of $\tilde{\tau} = S$ orthogonal pilot signals, but in
general we have $P$ pilots. \footnote{In the ideal case, for a large number of
users $K$, we would expect to have $P = S$, since $S$ is the maximum number
of mutually orthogonal vectors of length $S$, and will thus also give the
maximum number of simultaneous users that can be allocated pilots and served
by the base station.  However, in some cases the number of pilots used may
be smaller. We may for example have mobile users that require more frequent
pilots than static users (due to their channels having a shorter coherence
time), or we may assign multiple pilots within each coherence block to the
same user in order to improve the quality of the CSI obtained.} $P$ is thus
the maximum number of devices for which CSI can be obtained in each coherence
block. The pilot length $S$ and number of pilots $P$ are independent of the
total number of end devices that may be associated with the base station,
however, if, as in the scenarios we consider, the number of end devices $K$ is
greater than $P$, user scheduling is required to assign users to coherence
blocks, with no more than $P$ users active in each block.

Each end device $k \in \mathcal K$ has a coefficient $\beta(k)$ describing the
large scale effects on the device's channel. We further denote the uplink SNR by
$\hat{\rho}$ and the downlink SNR by $\check{\rho}$. These two parameters are as
defined in \cite{marzetta2016fundamentals}, Section 2.1.8, that is, they can be
interpreted as SNRs when the median of $\beta$ is 1.0, but in general $\rho$
scales with $\beta$. For each end device, we also have the mean-square channel
estimate, $\gamma(k)$. With perfect CSI, we have $\gamma(k) = \beta(k)$,
otherwise $\gamma(k)$ is given by \[\gamma(k) = \frac{S \hat{\rho} \beta(k)^2}{1
+ S \hat{\rho} \beta(k)}\] for all $k \in \mathcal K$.

On the uplink, each device $k \in \mathcal K$ transmits with a power control
coefficient $\hat\eta_k$, $0 \leq \hat\eta_k \leq 1$, where a power control
coefficient of $0$ indicates the device does not transmit at all, while $1$
indicates the device transmits with full power. On the downlink, the power control
coefficient $\check\eta_k$ indicates the power that the base station allocates
to transmission to device $k$. The sum of the downlink power control coefficients
gives the total normalized power with which the base station transmits, and so
the $\check\eta_k$ must sum to at most $1$, indicating full transmission power
from the base station.

We will consider two precoding schemes for the base station: maximum ratio
combining and zero forcing. In maximum ratio combining, we seek to maximize the
power of each device signal at that device (downlink), or when recovering the
received signal (uplink). For zero forcing, we instead seek to produce nulls in
the channel at devices other than the relevant one, thus creating zero
interference between the device signals if we have perfect CSI. However, there
is still interference in the case of imperfect CSI. The effective SINR for
the uplink and downlink for each of the two precoding schemes is shown in Table
\ref{tab:effective_SINR}. For a more complete discussion of the precoding
schemes, as well as the derivation of these expressions, see
\cite{marzetta2016fundamentals}, Chapter 3.

Note that the expressions given in the table apply when all $K$ devices are
active simultaneously, and there are sufficient pilots for each device to be
assigned one. For fewer active devices, the expressions need to be adjusted
accordingly, as we will in our optimization formulations in Section
\ref{sec:problem-solution}. Specifically, the summations in the denominators
should be taken only over the set of active users, rather than all $k \in
\mathcal K$, and the term $(M - K)$ that appears in the numerator of the zero
forcing expressions should instead become $(M - L)$, where $L$ is the number of
active users. In this paper, we do not consider pilot reuse --- that is,
multiple end devices using the same pilot in the same coherence block --- nor
the resulting pilot contamination.

\begin{table}
    \begin{center}
        \def\arraystretch{3}
        \begin{tabular}{|c|c|}
            \hline
            Maximum Ratio Combining & Zero Forcing \\
            \hline
             \(\displaystyle \frac{M\hat{\rho}\gamma(k)\hat\eta_k}{1 + \hat{\rho}
        \sum_{k' \in \mathcal K} \beta(k')\hat\eta_{k'}}\) &
            \(\displaystyle \frac{(M - K)\hat{\rho}\gamma(k)\hat\eta_k}{1 + \hat{\rho}\sum_{k' \in
            \mathcal K} \left(\beta(k') - \gamma(k')\right)\hat\eta_{k'}}\) \\[2ex]
            \hline
             \(\displaystyle \frac{M\check{\rho}\gamma(k)\check\eta_k}{1 +
        \check{\rho}\beta(k)\sum_{k' \in \mathcal K}\check\eta_{k'}}\) &
            \(\displaystyle \frac{(M - K)\check{\rho}\gamma(k)\check\eta_k}{1 +
                \check{\rho}\left(\beta(k) - \gamma(k)\right)\sum_{k' \in
            \mathcal K} \check\eta_{k'}}\) \\[2ex]
            \hline
        \end{tabular}
        \caption{Effective SINR expressions for precoding with maximum ratio
        combining and zero forcing, for uplink (top) and downlink (bottom)
        transmission (from \cite{marzetta2016fundamentals}, Section 3.4).}
        \label{tab:effective_SINR}
    \end{center}
\end{table}

\subsection{Traffic Model}
We now seek to schedule the traffic demands of the $K$ devices in as few
coherence blocks as possible. For each device $k \in \mathcal K$, we define
$\hat{h}(k)$ as the device's uplink demand, and $\check{h}(k)$ as the device's
downlink demand. The demands are expressed as a number of coherence blocks, that
is, to satisfy its uplink demand, device $k$ must transmit during the entire
uplink phase (of $\hat{\tau}$ channel uses) in $\hat{h}(k)$ coherence blocks,
and similarly for the downlink case. A device may transmit in both the uplink and
downlink phases of a given coherence block, but a device may not transmit in
either phase if it has not been allocated a pilot during that coherence block.
In order to transmit successfully, a device $k$ must also have an SINR in the
relevant transmission phase larger than or equal to its SINR threshold, $\mu(k)$.

Different devices may have different SINR thresholds; this means that the traffic
volume for devices in absolute terms (i.e. traffic volume in bytes) may not be
equal, even when devices have equal demands, as they may transmit at different
rates (achieved using different MCSs) during their assigned coherence blocks.
More precisely, if a device $k$ has $H(k)$ bytes of data to transmit, and
chooses an SINR threshold $\mu(k)$ allowing for a data rate of $m(\mu(k))$ bytes
per coherence block, then $k$'s traffic demand in coherence blocks is given by
$\frac{H(k)}{m(\mu(k))}$. In this way, devices can have both individual data volumes
and QoS requirements (specifically data rates).

The device demands could represent either the current traffic queued at the
devices, or recurring demands induced by periodic traffic as is typical in
sensor network monitoring scenarios. In the former case, information about the
device demands needs to be updated regularly at the base station, which places
more stringent constraints on the time needed to schedule the devices'
transmissions. With recurring demands, on the other hand, a schedule can be
established once and then used for a long time, making longer scheduling times
more feasible. Varying demands also incur a signaling overhead in order to
transmit information about device demands to the base station. However, this
overhead can be quite small, for example a single value representing the current
queue length, or may even be avoided by predicting traffic demands at the base
station. This is particularly feasible if the device traffic is not delay
sensitive, as it allows longer scheduling windows and/or the possibility to not
serve a device demand fully in a given scheduling window in case of traffic
prediction error.

\subsection{Device Scheduling}
We define a \emph{frame} as the set of coherence blocks required to meet all
device demands in the current scheduling window. We then seek to minimize the
frame, that is, find the least number of coherence blocks required for all
devices to send and receive all their traffic. Minimizing the frame thus
maximizes the throughput for contiguous frames, or, alternatively, maximizes the
sleep time of devices if there is a delay (sleep cycle) between successive
frames. Minimizing the frame also facilitates network slicing, an important
component of the 5G architecture \cite{foukas2017network}, by freeing more
resources to be used by other slices. The frame minimization problem consists of
allocating devices to coherence blocks in which they will transmit.

A set of devices that can successfully transmit and/or receive together during a
coherence block form a compatible set. In each c-set, each device may take the
role of a transmitter, a receiver, or both. If a node is a transmitter, it
transmits data during the uplink phase of the coherence block; if a node is a
receiver, it receives data during the downlink phase of the coherence block. If
a node has both roles, it is active in both phases. Each node in the c-set must
have a pilot assigned to it in order to transmit or receive during a given
coherence block, and as such, the number of devices in a c-set is bounded from
above by $P$, regardless of the achievable SINRs when transmitting or receiving
simultaneously.

This definition of a c-set differs to that used in wireless mesh networks (see
Section \ref{sec:related_work}) in three ways. Firstly, the expressions used
here for the SINRs are different, as they are derived from massive MIMO channel
models and precoding schemes, rather than the models for single omnidirectional
antennas typically used in mesh networks. Secondly, the number of available
pilots limits the number of nodes that may be placed in the same c-set. No more
nodes may be added to a c-set once all pilots are assigned, even if all nodes'
SINR conditions would be met. For a given number of pilots, this reduces the
number of possible c-set solutions, thus reducing the complexity of c-set
generation, which we will apply in order to solve the frame minimization
problem. Finally, in mesh network c-sets, a node cannot be both a transmitter
and receiver, whereas this is possible in massive MIMO c-sets; this is a
consequence of the massive MIMO coherence block structure, which exploits
channel reciprocity (in the TDD case) to use uplink pilots for both uplink and
downlink channel estimation.

The family of all possible c-sets is denoted by $\mathcal C$. The family of c-sets
in which device $k \in \mathcal K$ is a transmitter is denoted by $\hat{\mathcal
C}(k)$, and the family of c-sets in which device $k$ is a receiver is denoted by
$\check{\mathcal C}(k)$. Note that $\hat{\mathcal C}(k)$ and $\check{\mathcal
C}(k)$ are not necessarily disjoint. For a given c-set $c \in \mathcal C$,
$t(c)$ denotes the set of nodes that transmit in the uplink phase in coherence
blocks allocated to $c$, and $r(c)$ denotes the set of nodes that transmit in
the downlink phase in coherence blocks allocated to $c$. Again, $t(c)$ and
$r(c)$ are in general not disjoint.

\section{Frame Minimization Problem and Solution Approach} \label{sec:problem-solution}

In this section we will formulate the main problem of this paper and describe a
suitable approach for its optimization.

\subsection{Problem Formulation}

The optimization problem studied in this paper --- frame minimization --- is formulated as the following integer
programming (IP) problem. The problem will be referred to as the \emph{main problem} and denoted by MP/IP.
\begin{subequations} \label{form:frame}
    \begin{align}
        \min \quad & \sum_{c \in \mathcal C} T_c \label{objective-c-sets} \\
	    \quad & \sum_{c \in \hat{\mathcal C}(k)} T_c \ge \hat{h}(k), & k \in \mathcal K \label{c-set_uplink} \\
	    \quad & \sum_{c \in \check{\mathcal C}(k)} T_c \ge \check{h}(k), & k \in \mathcal K \label{c-set_downlink} \\
              & T_c \in \mathbb Z_+, & c \in \mathcal C. \label{c-integrality}
    \end{align}
\end{subequations}
Here, $\mathcal C$ is the family of all c-sets and $T_c$ are (integer) decision variables indicating the number of coherence blocks in
which the set of active devices (and their roles) is given by c-set $c \in
\mathcal C$. The objective \eqref{objective-c-sets} then seeks to minimize the
total number of coherence blocks used. Constraints \eqref{c-set_uplink} ensure,
for each device $k \in \mathcal K$, that the number of coherence blocks in which
the device will be active during the uplink phase of the block will be sufficient
to meet the device's uplink demand. Constraints \eqref{c-set_downlink} are similar,
but for the downlink demands.

Observe that in formulation \eqref{form:frame}, as in the remainder of this
paper, we follow the notational convention that variables are indexed by
subscripts (like index $c$ in variable $T_c$), and parameters (constants) by
round brackets (like index $k$ in parameter $\hat h(k)$). This convention, used
for example in \cite{korte-vygen12}, makes it easier to distinguish variables
from parameters in problem formulations.

\subsection{Solution Approach}\label{subsec:solution-algorithm}

In the solution approach presented below we will consider restricted versions of
MP/IP, denoted by MP/IP($\bar{\mathcal C}$), where only c-sets from a predefined
subfamily $\bar {\mathcal C}$ of the family $ \mathcal C$ (of all c-sets) are
allowed. In such a case the list of variables in \eqref{form:frame} is
restricted to $T_c, \, c \in \bar{\mathcal C}$, and the families $\hat{\mathcal
C}(k), \check{\mathcal C}(k), \, k \in \mathcal K,$ are restricted accordingly.
Below, the linear relaxation of MP/IP($\bar{\mathcal C}$) (where the decision
variables $T_c, \, c \in \bar{\mathcal C},$ are continuous rather than integer)
will be denoted by MP/LR($\bar{\mathcal C}$). Note that with this notation,
MP/IP($\mathcal C$) denotes the main problem MP/IP (i.e., formulation
\eqref{form:frame}), and MP/LR($\mathcal C$) its linear relaxation. The latter
formulation will be simply referred to as MP/LR.

Since the total number of c-sets (and hence the number of variables $T_c$) grows
exponentially with the number of devices and pilots, the MP/IP formulation (and,
for that matter, the MP/LR($\mathcal C$) formulation also) becomes non-compact.
Hence, it is not feasible in general to solve the frame minimization problem
\eqref{form:frame} using all possible c-sets. Instead, we apply column
generation \cite{ahuja1993}.

The approach is to start (in Phase~1) by solving the linear relaxation
MP/LR($\mathcal C$) using column generation --- in our context column generation
is referred to as \emph{c-set generation} \cite{Pioro2011,PTC-2018} since in
formulation \eqref{form:frame} columns, i.e., variables, correspond to c-sets
--- and then (in Phase~2) solve MP/IP($\mathcal C^*$) restricted to the family
$\mathcal C^*$ of the c-sets resulting from Phase~1.

\subsubsection{Phase~1. Solving the linear relaxation of the main problem by c-set generation} \label{susubsec: Phase 1}

Consider the linear relaxation MP/LR($\bar{\mathcal C}$) for a given subfamily (list) $\bar{\mathcal C}$ of the family of all c-sets $\mathcal C$, i.e., the following linear programming problem formulation:
\begin{subequations} \label{form:frame-LR}
    \begin{align}
        \min \quad & \sum_{c \in \bar{\mathcal C}} T_c \label{objective-c-sets-LR} \\
	[ \hat{\pi}_k \ge 0 ] \quad & \sum_{c \in \hat{\mathcal C}(k) \cap \bar{\mathcal C}} T_c \ge \hat{h}(k), & k
        \in \mathcal K \label{c-set_uplink-LR} \\
	[ \check{\pi}_k \ge 0 ] \quad & \sum_{c \in \check{\mathcal C}(k) \cap \bar{\mathcal C}} T_c \ge \check{h}(k), & k
        \in \mathcal K \label{c-set_downlink-LR} \\
                                  & T_c \in \mathbb R_+, & c \in \bar{\mathcal C}. \label{c-continuity-LR}
    \end{align}
\end{subequations}
The above problem will be called the \emph{primal problem}.

Now, using the (dual) variables specified in square brackets to the left of constraints \eqref{c-set_uplink-LR} and \eqref{c-set_downlink-LR}, we form the dual to the (linear programming) primal problem \eqref{form:frame-LR} to give the following linear programming problem formulation \cite{lasdon1970,minoux1986}:
\begin{subequations} \label{dual}
    \begin{align}
	\max \quad & \sum_{k \in \mathcal K} \left( \hat{\pi}_k \hat{h}(k) + \check{\pi}_k \check{h}(k) \right) \label{objective-dual} \\
    & \sum_{k \in t(c)} \hat{\pi}_k + \sum_{k \in r(c)} \check{\pi}_k  \leq 1, & c \in \bar{\mathcal C} \label{constraint-dual} \\
    & \hat{\pi}_k, \check{\pi}_k \in \mathbb R_+, & k \in \mathcal K. \label{variables-dual}
    \end{align}
\end{subequations}
(Recall that $t(c)$ ($r(c)$) denotes the set of devices $k \in \mathcal K$ that transmit (receive) in compatible set $c$; see Table~\ref{tab:notation}.) This \emph{dual problem} will be denoted by DP($\bar{\mathcal C}$) and  referred to as the \emph{master problem} in the column generation algorithm formulated below.

Consider a dual optimal solution $\pi^* = ((\hat \pi^*_k, \check \pi^*_k), k \in
\mathcal K)$, and suppose there exists a c-set, $c'$, say, outside the current
list $\bar{\mathcal C}$ with $\sum_{k \in t(c')} \hat{\pi}^*_k + \sum_{k \in
r(c')} \check{\pi}^*_k > 1$. When $c'$ is added to the list ($\bar{\mathcal C}
:= \bar{\mathcal C} \cup \{ c' \}$) then the new dual has one more constraint
\eqref{constraint-dual} that corresponds to $c'$, and this particular constraint
is broken by the current optimal solution $\pi^*$. This means that the new dual
polytope (for the updated c-set list $\bar{\mathcal C}$) determined by
conditions \eqref{constraint-dual}-\eqref{variables-dual} is a proper subset of
the previous dual polytope, as the current optimal dual solution is cut off by
the new dual constraint. Therefore, in the updated dual the maximum of
\eqref{objective-dual} cannot be greater than the previous maximum, and in fact
in most cases it will be decreased. Thus, taking into account that the maximum
of the dual problem is equal to the minimum of the primal problem (this is a
general property of convex problems, and thus linear programming problems like
MP/LR, called the \emph{strong duality property} in optimization theory
\cite{lasdon1970,minoux1986}), adding the c-set $c'$ will usually decrease the
frame length. On the other hand, if no such new c-set $c'$ exists, the final
dual (and primal) solution thus obtained is optimal even if $\bar{\mathcal C}$
were to be extended to the list of all possible c-sets. The reason for this is
that $\pi^*$ fulfills constraint \eqref{constraint-dual} for \emph{all} $c \in
\mathcal C$. 

The problem of finding a c-set $c'$ that maximizes the quantity
\begin{equation} \label{eq:PP-B}
B(c;\pi^*) = \sum_{k \in t(c)} \hat{\pi}^*_k + \sum_{k \in r(c)} \check{\pi}^*_k
\end{equation}
over the family $\mathcal C$ (i.e, the family of all c-sets) will be called the \emph{pricing problem} and denoted by PP($\pi^*$). The pricing problem, a crucial element of the column generation algorithm, will be discussed in detail in Section~\ref{subsec:PP}.

The above observations lead to the following iterative \emph{c-set generation algorithm} that solves the linear relaxation of the main problem MP/IP.
\vspace{0.15cm}

\noindent \textbf{CG (c-set generation) algorithm}
\begin{description}
\item [Step 1:] Define initial $\mathcal C^0$ and let $\bar{\mathcal C} := \mathcal C^0$.
\item [Step 2:] Solve the master problem DP($\bar{\mathcal C}$) defined in \eqref{dual}; let $\pi^* = ((\hat \pi^*_k, \check \pi^*_k), k \in \mathcal K)$ be the resulting optimal dual solution.
\item [Step 3:] Solve the pricing problem PP($\pi^*$); let $c'$ be a c-set that maximizes the quantity $B(c;\pi^*)$ defined in \eqref{eq:PP-B} over $c \in \mathcal C$.
\item [Step 4:] If $B(c') > 1$, then $\bar{\mathcal C} := \bar{\mathcal C} \cup \{ c' \}$ and go to Step~2.
\item [Step 5:] $\mathcal C^* := \bar{\mathcal C}$; solve the primal problem MP/LR($\mathcal C^*$) defined in \eqref{form:frame-LR}, and stop.
\end{description}
\vspace{0.15cm}
When the CG algorithm stops, the primal solution calculated in Step~5 solves the linear relaxation, MP/LR($\mathcal C$), of the main problem MP/IP. Certainly, for the CG algorithm to start properly, the predefined initial list of c-sets, $\mathcal C^0$, appearing in Step~1 must assure feasibility of MP/LR($\mathcal C^0$). Observe also that the pricing problem in Step~3 generates each new c-set $c'$ (if any), whose constraint \eqref{constraint-dual} is \emph{maximally broken} by $\pi^*$. More information on the CG algorithm will be given in Section~\ref{subsec:full-LR}.

\subsubsection{Phase~2. Solving the main problem} \label{sububsec: Phase 2}

After generating the c-set list $\mathcal C^*$, the main problem \eqref{form:frame} restricted to $\mathcal C^*$  (denoted by MP/IP($\mathcal C^*$)), is
solved by the branch-and-bound (B\&B) algorithm \cite{pioro2012network}. Note that the integer solution, $t^*_c,\, c \in \mathcal C^*$, obtained thereby may in general be suboptimal, since there can exist c-sets that
are not necessary to solve the linear relaxation \eqref{form:frame-LR} but are required for achieving the optimum of the MP/IP (where all c-sets are considered).
We will return to this issue in Section~\ref{subsec:main-problem}.

\subsection{Pricing Problems} \label{subsec:PP}

The form of the pricing problem will depend on the precoding method used by the
base station. Below, we give pricing problems for both maximum ratio combining and
zero forcing. Both of these problems require variable multiplications (so-called bi-linearities) that need
to be resolved by adding auxiliary variables and constraints. We however omit
these in the following. In the appendix, we describe the necessary auxiliary
variables and accompanying constraints to render the pricing problems as
proper mixed-integer programming formulations. Also, discussion of
computational efficiency issues of the pricing problems will be deferred to Section~\ref{subsec:eff-pricing-problem}.

\subsubsection{Maximum Ratio Combining (MRC)} \label{subsubsec:maximum-ratio-combining}
The pricing problem for MRC can be formulated as follows:
\begin{subequations} \label{PP-MRC}
    \begin{align}
	\max \quad & \sum_{k \in \mathcal K} \left(\hat{u}_k \hat{\pi}^*_k +
	\check {u}_k \check{\pi}^*_k \right)
	\label{objective-PP-MRC} \\
	& \Delta \left(1 - \hat{u}_k \right) + M \hat{\rho} \gamma(k)
	\hat{\eta}_k \geq \nonumber \\
	& \qquad \mu(k) \Big( 1 + \hat{\rho} \sum_{k' \in \mathcal
	K} \beta(k') \hat{\eta}_{k'} \hat{u}_{k'} \Big), & k \in \mathcal K \label{uplink-SINR-MRC} \\
	& \Delta \left(1 - \check{u}_k \right) + M \check{\rho} \gamma(k)
	\check{\eta}_k \geq \nonumber \\
	& \qquad \mu(k) \Big( 1 + \check{\rho} \beta(k)
	\sum_{k' \in \mathcal K} \check{\eta}_{k'} \check{u}_{k'} \Big), & k \in \mathcal K \label{downlink-SINR-MRC} \\
	& u_k \geq \hat{u}_k; \,
	u_k \geq \check{u}_k; \,
	u_k \leq \hat{u}_k + \check{u}_k, & k \in \mathcal K \label{x-MRC} \\
	& \sum_{k \in \mathcal K} u_k \leq P \label{pilots-MRC} \\
    & \hat{u}_k, \check{u}_k, u_k \in \mathbb B, & k \in \mathcal K \label{pilots-variables}\\
    & \eta \in \Theta, \label{power-variables}
    \end{align}
\end{subequations}
where the decision variables $u = (\hat{u}_k, \check{u}_k, u_k, \, k \in \mathcal K)$ (called pilot variables) and $\eta = (\hat{\eta}_k, \check{\eta}_k, \, k \in \mathcal K)$ (power control variables) are listed in \eqref{pilots-variables} and \eqref{power-variables}, respectively. The constant $\Delta$ appearing in constraints \eqref{uplink-SINR-MRC}-\eqref{downlink-SINR-MRC} is defined as $\Delta = N\left(K\hat{\rho}B + 1\right)$, for $B = \mathrm{max}\{\beta(k) : k \in \mathcal K\}$ and $N = \mathrm{max}\{\mu(k) : k \in \mathcal K\}$. Note that in the above formulation the constraint on the power control variables is not explicit and involves a set $\Theta$ of allowable vectors $\eta$. This set, a parameter of the pricing problem, will be defined in Section~\ref{subsec:power-control} for selected power control schemes. Note also the quantities $\hat{\pi}^*_k$ and $\check{\pi}^*_k$ (optimal values of the dual variables) appearing in objective \eqref{objective-PP-MRC} are basic parameters of the pricing problem.

The binary decision variable $u_k$ will be equal to 1 if, and only if, end device $k
\in \mathcal K$ is to be included in the new c-set. Similarly, $\hat{u}_k$ and
$\check{u}_k$ indicate whether end device $k$ is to be a transmitter and/or
receiver, respectively, in the new c-set. The objective \eqref{objective-PP-MRC}
selects a new c-set that will maximally violate the corresponding constraint
\eqref{constraint-dual} in the dual problem. Constraints \eqref{uplink-SINR-MRC}
and \eqref{downlink-SINR-MRC} ensure the SINR thresholds are met for each device
for the uplink and downlink phases, respectively, and are based on the
expressions given in Table \ref{tab:effective_SINR}.

Constraints \eqref{x-MRC} ensure that a device $k$ is included in the c-set if it is set as a transmitter or receiver, and that the device is not included in the c-set if it is inactive in both the uplink and downlink phases. Finally, constraint \eqref{pilots-MRC} ensures that the c-set does not contain a greater number of devices than there are pilots available to assign to them in a given coherence block.
Clearly, if $u^*$ is an optimal vector $u$, then the optimal c-set $c'$ that solves the pricing problem is determined as follows: $c' = \{ k \in \mathcal K: \, u^*_k =1 \}$, with $t(c') = \{ k \in \mathcal K: \, \hat{u}^*_k =1 \}$ and $r(c') = \{ k \in \mathcal K: \, \check{u}^*_k =1 \}$.

\subsubsection{Zero Forcing (ZF)} \label{subsubsec:zero-forcing}
The pricing problem for ZF is similar to that for MRC, but instead uses the zero forcing effective SINR expressions.
\begin{subequations} \label{PP-ZF}
    \begin{align}
	\max \quad & \sum_{k \in \mathcal K} \left(\hat{u}_k \hat{\pi}^*_k +
	\check{u}_k \check{\pi}^*_k \right)
	\label{objective-PP-ZF} \\
	& \Delta \left(1 - \hat{u}_k\right) + \left( M - \hat L \right) \hat{\rho} \gamma(k)
	\hat{\eta}_k \geq \nonumber \\
	& \qquad \mu(k) \Big( 1 + \hat{\rho} \sum_{k' \in \mathcal K} \left( \beta(k') - \gamma(k') \right)
	\hat{\eta}_{k'} \hat{u}_{k'} \Big), \hspace{-6em}\nonumber \\
	& &  k \in \mathcal K \label{uplink-SINR-ZF} \\
	& \Delta \left(1 - \check{u}_k \right) + \left( M - \check L \right) \check{\rho}
	\gamma(k) \check{\eta}_k \geq  \nonumber \\
	& \qquad \mu(k) \Big( 1 + \check{\rho} \left( \beta(k) - \gamma(k) \right)
    \sum_{k' \in \mathcal K} \check{\eta}_{k'} \check{u}_{k'} \Big),
	\hspace{-6em}\nonumber \\
	& & k \in \mathcal K \label{downlink-SINR-ZF} \\
	& \hat L = \sum_{k \in \mathcal K} \hat{u}_k &  \label{L0-ZF} \\
	& \check L = \sum_{k \in \mathcal K} \check{u}_k &  \label{L1-ZF} \\
	& u_k \geq \hat{u}_k; \, u_k \geq \check{u}_k; \, u_k \leq \hat{u}_k + \check{u}_k, & k \in \mathcal K \label{x-inactive-ZF} \\
	& \sum_{k \in \mathcal K} u_k \leq P, \label{pilots-ZF}\\
    & \hat L, \check L \in \mathbb Z_+ \label{L01-ZF-var}\\
    & \hat{u}_k, \check{u}_k, u_k \in \mathbb B, & k \in \mathcal K \label{pilots-variables-ZF}\\
    & \eta \in \Theta. \label{power-variables-ZF}
    \end{align}
\end{subequations}
Integer variables $\hat L$ and $\check L$ are introduced and set to the number of
active nodes in the uplink and downlink phases, respectively, via constraints
\eqref{L0-ZF} and \eqref{L1-ZF}. This is needed for the SINR expressions in
constraints \eqref{uplink-SINR-ZF} and \eqref{downlink-SINR-ZF}, in order to
take the number of active users rather than the total number of users on the
left hand side.

\subsection{Adding Power Control Optimization to Pricing} \label{subsec:power-control}

The pricing problems given above are incomplete since the constraints on the power control
coefficients $\hat\eta_k$ and $\check\eta_k$ are not explicitly defined. There are a number of different possible power control schemes that may be used. Here, we
will consider three of them to compare, where each of these schemes may optionally include power control on the uplink or not. In all the considered cases power control optimization is achieved by specifying the set $\Theta$ (appearing in constraints \eqref{power-variables} and \eqref{power-variables-ZF}) by means of explicit constraints on variables $\eta$.

\subsubsection{Joint Power and C-Set Composition Optimization} \label{subsubsec:power-1}

The first scheme is to jointly optimize the power control coefficients $\eta$ and the c-set composition. This requires the constraints
\begin{subequations}
    \begin{align}
	& \hat{\eta}_k \leq 1, & k \in \mathcal K \label{device_max_power} \\
	& \sum_{k \in \mathcal K} \check{\eta}_k \leq 1 \label{BS_max_power} \\
    & \hat{\eta}_k, \check{\eta}_k \in \mathbb R_+, & k \in \mathcal K
    \end{align}
\end{subequations}
to be incorporated into both the pricing problems \eqref{PP-MRC} (for MRC) and
\eqref{PP-ZF} (for ZF) instead of \eqref{power-variables} and
\eqref{power-variables-ZF}, respectively. The two constraints simply ensure that
the devices and the base station, respectively, do not exceed their maximum transmission power.

\subsubsection{Fair Power Control Optimization} \label{subsubsec:power-2}

The second power control scheme is (max-min) fair power control, as defined in
\cite{marzetta2016fundamentals}, Section 5.3. In this scheme, nodes adjust their
transmission power so as to maximize the minimum SINR of any node. It can be proven
that this always results in all nodes achieving a single, common SINR --- see
\cite{marzetta2016fundamentals}, Section 5.3.1. Nodes adjust their power
according to their relative channel quality, such that the node(s) with the
worst channel(s) will transmit at full power, while nodes with good channels
reduce their transmission power. For this scheme, in the case of MRC we need to
replace \eqref{power-variables} with the following constraints.
\begin{subequations}\label{MRC_fair}
    \begin{align}
	& \hat{\eta}_k = \frac{\varphi}{\gamma(k)}, & k \in \mathcal K
	\label{uplink-power-MRC} \\
	& \varphi \leq \gamma(k) \hat{u}_k + (1 - \hat{u}_k)\Gamma, & k \in \mathcal K \label{phi-MRC-l} \\
	& \varphi \geq \sum_{k \in \mathcal K} \gamma(k) z_k \label{phi-MRC-g} \\
	& \sum_{k \in \mathcal K} z_k \hat{u}_k \leq 1 \label{z-1-MRC} \\
	& \sum_{k' \in \mathcal K} z_{k'} \hat{u}_{k'} \geq \hat{u}_k, & k \in
	\mathcal K \label{z-2-MRC} \\
	& \check{\eta}_k \check{\rho} \gamma(k) \left( \frac{1}{\check{\rho}} \sum_{k' \in \mathcal
	    K} \frac{\check{u}_{k'}}{\gamma(k')} + \sum_{k' \in \mathcal K}
	    \frac{\beta(k') \check{u}_{k'}}{\gamma(k')}\right) =  \nonumber \\
	& \qquad\qquad\qquad \left(1 + \check{\rho}
	\beta(k) \right) \check{u}_k, & k \in \mathcal K, \label{downlink-power-MRC} \\
    & z_k \in \mathbb B, & k \in \mathcal K \\
    & \hat{\eta}_k, \check{\eta}_k, \varphi \in \mathbb R_+, & k \in \mathcal K,
    \end{align}
\end{subequations}
where
$\Gamma = \mathrm{max}\{\gamma(k) : k \in \mathcal K\}$. Constraints
\eqref{phi-MRC-l}--\eqref{z-2-MRC} select the lowest $\gamma(k)$ of any node
active in the uplink phase to be the numerator in constraint
\eqref{uplink-power-MRC}, with constraint \eqref{phi-MRC-l} providing an upper
bound,  and constraints \eqref{phi-MRC-g}--\eqref{z-2-MRC} providing a lower
bound by selecting one active device, if there are any. If no devices are active
in the uplink phase, $\varphi$ will be zero. Constraints \eqref{downlink-power-MRC}
are taken from the fair power control expressions derived in
\cite{marzetta2016fundamentals}, Section 5.3.1.

For fair power control in the case of ZF, we similarly need to add the above constraints to the ZF
pricing problem \eqref{PP-ZF}, except that the following is used instead of
constraint \eqref{downlink-power-MRC}.
\begin{align}
	& \check{\eta}_k \check{\rho} \gamma(k) \left( \frac{1}{\check{\rho}} \sum_{k' \in \mathcal
	    K} \frac{\check{u}_{k'}}{\gamma(k')} + \sum_{k' \in \mathcal K}
	\frac{\left( \beta(k') - \gamma(k') \right)
	\check{u}_{k'}}{\gamma(k')}\right) = \nonumber \\
	& \qquad\qquad\qquad \left( 1 + \check{\rho}
    \left( \beta(k) - \gamma(k) \right) \right) \check{u}_k,
	\qquad\quad k \in
	\mathcal K. \label{ZF_fair}
\end{align}
Here, the difference to the MRC case is that $\beta(k)$ is replaced by
$\left( \beta(k) - \gamma(k) \right)$.

\subsubsection{Static Power Control Optimization} \label{subsubsec:power-3}

Since the above power control schemes require changing the power control
coefficients each time a new c-set becomes active, they add a significant
signaling overhead to inform the end devices of the values of the power control
coefficients they should use. At the base station, new power control
coefficients must be calculated for each coherence block, or, alternatively,
stored for each c-set. To alleviate these problems, we may instead use a third
scheme, which we will call static power control, in which the variables
$\hat{\eta}_k$ and $\check{\eta}_k$ are removed from the power control
constraints by replacing each of them with $\frac{\mathrm{min}\{\gamma(k') : k'
\in \mathcal K\}}{\gamma(k)}$, thus calculating power control coefficients over
the entire set of $K$ devices. The $\hat\eta_k$ and $\check\eta_k$ then become
constant parameters instead of decision variables.

It is also possible to not perform any power control on the uplink. This is
simpler to implement on resource-constrained end devices. In this case, all
$\hat{\eta}_k$ are set to 1 for each of the above schemes. However, the power
control constraints for the downlink must still be determined using one of the
other methods. Without power control on the uplink, near-far effects must
instead be mitigated by the assignment of devices to c-sets.

\section{Complexity and Efficiency of the Solution Approach} \label{sec:efficiency}

The solution approach presented in Section~\ref{sec:problem-solution} is, by the
very nature of the frame minimization problem MP/IP, quite complicated. Although
we cannot definitely say that MP/IP is not polynomial, we may expect that its
linear relaxation MP/LR($\mathcal C$) is already $\mathcal{NP}$-hard. Reasons
for this include that the linear relaxation in question is non-compact (it has
an exponential number of variables); that the pricing problems required in the
CG algorithm (formulated in Section~\ref{subsec:PP}) are very similar to the
$\mathcal{NP}$-hard maximum clique problem \cite{korte-vygen12}; and also
that analogous pricing problems, for generating c-sets in wireless mesh
networks, are $\mathcal{NP}$-hard (see \cite{Goussevskaia2007,PTC-2018}). Thus,
assuming that the pricing problems used in this paper are also
$\mathcal{NP}$-hard, we can expect, by polynomial equivalence of separation and
optimization \cite{nemhauser-wolsey,nptz2013}, polynomiality of MP/LR($\mathcal
C$) to be unlikely. Bearing this is mind, finding a compact formulation for
MP/LR($\mathcal C$) is also unlikely since compact linear problems are solvable
in polynomial time \cite{nemhauser-wolsey}.

In the following we will discuss the efficiency issues related to our proposed
approach and comment on its heuristic aspects. The discussion of this section
will be further illustrated by the results presented in
Section~\ref{sec:heuristic_results}.

\subsection{Solving the Linear Relaxation} \label{subsec:full-LR}

As already mentioned, the predefined initial list of c-sets, $\mathcal C^0$,
appearing in Step~1 of the CG algorithm must assure feasibility of
MP/LR($\mathcal C^0$), which means that each $k \in \mathcal K$ must be
contained in at least one c-set in $\mathcal C^0 \cap \hat{\mathcal C}$ and in
at least one c-set $\mathcal C^0 \cap \check{\mathcal C}$, unless $k$'s uplink or
downlink demand (respectively) is zero. An example of such a $\mathcal C^0$ is the family composed of all c-sets in which only a single device transmits and receives in the coherence block; this is always a valid c-set since there will be no interference and only one pilot signal is needed for the single device. Clearly, the minimal frame length for this particular $\mathcal C^0$ is equal to $\sum_{k \in \mathcal K} \max \{ \hat h(k), \check h(k) \}$.

The final solution of the column-generation algorithm is always optimal for any
initial c-set list $\mathcal C^0$. Thus, the choice of a particular $\mathcal
C^0$ affects only the number of iterations of the CG algorithm, but usually to a
small extent. It turns out that the CG algorithm achieves a solution close to the
optimum of MP/LR($\mathcal C$) very quickly (in fewer than 20 iterations in our
study, see Section~\ref{subsubsec:exp-1-3}), no matter what the initial list
$\mathcal C^0$ is. In effect, most of the iterations are spent on decreasing
such a quickly found suboptimal solution to eventually reach the optimum which
is not significantly smaller. Hence, the choice of a particular initial c-set
list is not really important. This phenomenon is in fact typical for non-compact
linear programming problems solved by column generation; an illustrative example
can be found in \cite{PTC-2018}.

The CG algorithm generates the columns of the primal problem (corresponding to
the c-sets) using the dual \eqref{dual} as the master problem in the main loop.
This means that in every iteration a new constraint \eqref{constraint-dual}
corresponding to the c-set $c'$ found by the pricing problem is added to the
master. Observe that this is equivalent to classical column generation using the
simplex method \cite{ahuja1993}. If the primal problem MP/LR($\bar{\mathcal C}$)
is solved by the simplex algorithm, then when a new column (variable $T_{c'}$)
is added to \eqref{form:frame-LR} and this column replaces one of the current
basic variables, the local rate of decrease in the value of the primal objective
\eqref{objective-c-sets-LR} will be equal to $\sum_{k \in \hat{c}} (\hat \pi^*_k
+ \sum_{k \in \check{c}} \check \pi^*_k) - 1$ (in the simplex method this value
is called the reduced cost of variable $T_{c'}$), and thus maximal over all
c-sets. After column $c'$ enters the basis, the value of
\eqref{objective-c-sets-LR} will be decreased by $(1 - \sum_{k \in \hat{c}}
\hat{\pi}^*_k + \sum_{k \in \check{c}} \check{\pi}^*_k ) t$, where $t$ is the
value assigned to variable $T_{c'}$ by the simplex pivot operation. If the
current basic solution happens to be degenerate then $t$ may have to stay at the
zero value, and in effect the objective function will not be decreased.
Nevertheless, adding variable $T_{c'}$ to the problem is necessary for the
simplex algorithm to proceed towards the optimal vertex solution. Certainly,
primal problem \eqref{form:frame-LR} could equally be used for the master problem (instead of the dual) since optimal dual variables $\pi^*$, the parameters of the pricing problem, can be straightforwardly calculated from the optimal primal simplex basis \cite{lasdon1970}.

It should be noted that the column generation method (and hence the CG algorithm), just like the simplex algorithm, does not guarantee a polynomial
number of iterations to reach the optimum (what is guaranteed is a finite number of iterations). Fortunately, in practical applications (like ours) the simplex algorithm is efficient and requires a polynomial number of steps (typically proportional to the number of variables), and so does column generation. In any case, however, there is virtually no alternative to column generation when it comes to solving non-compact linear programs.

As already mentioned in Section~\ref{susubsec: Phase 1}, when the CG algorithm
terminates, the final c-set list, $\mathcal C^*$, will contain all c-sets
necessary to solve the full linear relaxation of the main problem. This means that any optimal solution of MP/LR($\mathcal C^*$) is optimal for
MP/LR($\mathcal C$), i.e., the linear relaxation of the main problem MP/IP formulated in \eqref{form:frame}. Moreover, in the final simplex solution of
MP/LR($\mathcal C^*$) the optimal vertex (optimal basic solution of the standard
form of the primal problem) will contain at most $2|\mathcal K|$ non-zero values
in $t^* = (t^*_c, \, c \in \mathcal C^*)$. This observation implies that the gap
between the optimal solution of MP/IP and its lower bound computed from
MP/LR($\mathcal C^*$) is not greater than $2|\mathcal K|$ because the vector
$\lceil t^* \rceil := (\lceil t^*_c \rceil, \, c \in \mathcal C^*)$ is a
feasible solution of the main problem MP/IP. In fact, in practice this gap can
be considerably smaller than $2|\mathcal K|$ because the actual number of
non-zero elements in $t^*$ is equal to $2|\mathcal K|$ minus the number of
non-binding constraints in \eqref{c-set_uplink-LR}-\eqref{c-set_downlink-LR} for
$t^*$. Moreover, if we assume that the fractional parts of $t^*_c$ are random
then the average gap will be equal to $\sum_{c \in \mathcal C^*} \big( \lceil
t^*_c \rceil - t^*_c \big) = \frac{K'}{2}$ (where $K'$ is the number of non-zero
elements in vector $t^*$). This means, that the quality (strength) of the linear
relaxation MP/LR($\mathcal C^*$) will be good when the optimal value of the
objective function \eqref{objective-c-sets-LR} is large compared with
$2|\mathcal K|$, which is the case when the demands $\hat h(d)$ and/or $\check
h(d)$ are large. Application scenarios where this would occur include monitoring
or factory control networks, where traffic consists of regular updates with
fixed sizes, so that traffic can be reliably predicted over a long period of
time. Fortunately, it happens that even if this is not the case, the
quality of the lower bound is very good, as illustrated in
Section~\ref{sec:heuristic_results}.

\subsection{Solving the Pricing Problems }\label{subsec:eff-pricing-problem}

The pricing problems formulated in Section~\ref{subsec:PP} and used in
successive iterations of the CG algorithm are mixed-integer programming problems
with the number of binary decision variables proportional to $|\mathcal K|$
(including additional variables to eliminate the bi-linearities, see the
appendix). This is a reasonable number and, as it turns out, treating the
pricing problems directly with the CPLEX mixed-integer programming solver, which
applies the branch-and-bound algorithm, is sufficiently efficient
compared to the remaining elements (i.e., the main problem, and the master
problem in the CG algorithm) of the considered two-phase approach. Thus, we did
not attempt to apply additional integer programming means, like introducing
extra cuts to strengthen the linear relaxations of the pricing problems or even
using the branch-and-cut (B\&C) version of the B\&B algorithm (see
\cite{pioro2012network}) to improve the PP solution process efficiency.

\subsection{Solving the Main Problem }\label{subsec:main-problem}

Solving the main problem \eqref{form:frame} for a given (limited) list
$\bar{\mathcal C}$ of c-sets requires a branch-and-bound algorithm
\cite{pioro2012network}. Such a B\&B algorithm generates a rooted binary tree
(called the B\&B tree) composed of nodes (called the B\&B nodes). Each such B\&B
node corresponds to a particular linear programming subproblem obtained from the
linear relaxation MP/LR($\bar{\mathcal C}$) by restricting the range of
variables $T_c, \, c \in \bar{\mathcal C},$ through the B\&B node-specific lower
and upper bounds. More precisely, extra constraints of the form $L(B,c) \le T_c
\le U(B,c), \, c \in \bar{\mathcal C},$ are added to MP/LR($\bar{\mathcal C}$)
in the subproblem of the B\&B node $B$, where $L(B,c),\, U(B,c)$ are the lower and
upper bounds on $T_c$, respectively, specific to B\&B node $B$. The
algorithm starts with visiting the root of the B\&B tree, where $L(B,c) = 0$ and
$U(B,c) = \infty$ for all $c \in \bar{\mathcal C}$. In general, when the
algorithm visits a B\&B node $B$, it solves the subproblem of $B$, selects
a fractional $t^*_{c'}$ in the resulting optimal solution $t^* =(t^*_c, \, c \in
\bar{\mathcal C})$, and (provided there is a fractional component in $t^*$)
creates two new B\&B nodes $B'$ and $B''$ that are called active nodes. In $B'$,
the constraint $L(B,c') \le T_{c'} \le U(B,c')$ is substituted by $L(B',c') \le
T_{c'} \le \lfloor t^*_{c'} \rfloor$, and in $B''$ by $\lceil t^*_{c'} \rceil
\le T_{c'} \le U(B'',c')$. Note that if vector $t^*$ is integer, a feasible
integer solution (i.e., a feasible solution of MP/IP($\bar{\mathcal C}$)) is
achieved and the new B\&B nodes are not created. Nor are new B\&B nodes created
when the optimal objective of the subproblem of $B$ (i.e., $\sum_{c \in
\bar{\mathcal C}} t^*_c$) is greater than or equal to the current best feasible
integer solution. After visiting a B\&B node $B$ the algorithm proceeds to one
of the other active B\&B nodes and $B$ becomes inactive. There are several reasonable ways of visiting the active nodes, among them depth-first search (of the B\&B tree).

In fact, to assure true optimality, problem MP/IP should be solved using the branch-and-price (B\&P) algorithm \cite{pioro2012network} instead of the price-and-branch (P\&B) algorithm that underlies the two phase approach described in Section~\ref{sec:problem-solution}. The basic difference between B\&P and P\&B is that in the latter the CG algorithm is invoked only once, at the root node of the B\&B tree, and then the linear subproblem solved at each of the subsequent B\&B nodes assumes the subfamily $\mathcal C^*$ computed at the root. B\&P in turn, would apply the CG algorithm at each B\&B node. This would consume excessive overall computational time already for medium size MIMO systems because of executing the CG algorithm at every B\&B node.

With P\&B the linear subproblems solved at the B\&B nodes (obtained, as
explained above, by adding constraints on the range of (continuous) variables
$T_c, \, c \in \mathcal C^*,$ to MP/LR($\mathcal C^*$)), are solved quickly by
the CPLEX linear programming solver, provided the number of c-sets in $\mathcal
C^*$ is reasonable (this is the case in the numerical examples considered in
Section~\ref{sec:experiments}). Moreover, the limited number of decision
variables $T_c, \, c \in \mathcal C^*,$ and good quality of the linear
relaxations (tight lower bounds on the objective value of the corresponding
MP/IP($\mathcal C^*$)) make the P\&B solution process sufficiently efficient for
our purposes. Here we could also allow decreasing the family $\mathcal C^*$ used for solving MP/IP by letting the CG algorithm stop after relatively few iterations when the slope of the gain in the optimal value of the master is becoming flat and the current master solution is close to its final, optimal, value (see Section~\ref{subsubsec:exp-1-3}).

It is important that the optimal solutions obtained with our P\&B two-phase
approach, i.e., optimal solutions of MP/IP($\mathcal C^*$), are close to the
lower bound on MP/IP (obtained with MP/LR($\mathcal C$)), as shown in
Section~\ref{sec:heuristic_results}. It is also worth noticing that good quality
solutions of the main problem \eqref{form:frame} can be obtained by solving
MP/IP($\mathcal C'$), where $\mathcal C' = \{ c \in \mathcal C^*: \, t^*_c > 0
\}$, $\mathcal C^*$ is the family of c-sets obtained in Phase~1, and $t^*$ is
the final optimal solution of Phase~1, i.e., of MP/LR($\mathcal C^*$). This
observation is illustrated in Section~\ref{sec:heuristic_results}. 
As already mentioned, the number of the c-sets in family $\mathcal C'$ is not greater than $2K$ and can be much smaller than the number of the c-sets in family $\mathcal C^*$. Hence, the number of (integer) variables in MP/IP($\mathcal C'$) can be much smaller than in MP/IP($\mathcal C^*$).

To end this section we note that it is possible to write down a (compact)
mixed-integer problem formulation for MP/IP with the number of variables and
constraint polynomial in $K$ and the maximum frame length. In such a
formulation, the c-sets for the consecutive coherence blocks of the optimized
frame are specified explicitly by means of additional binary variables and
constraints (for each coherence block) in the way used in the pricing problems.
Such a formulation could be solved directly, using a mixed-integer programming
solver, for example the solver available in CPLEX. However, the formulation in
question would involve a number of binary variables (and constraints) that is far beyond the reach of IP solvers. Therefore, the c-set generation algorithm (where the pricing problem for finding the improving c-set in each iteration of the CG algorithm is used only once per iteration) applied in the proposed two-phase approach seems to be the only reasonable option for reaching good quality (near optimal) solutions of the main problem MP/IP.

\section{Numerical Study}\label{sec:experiments}
For our numerical study, formulations \eqref{form:frame}--\eqref{ZF_fair} were
implemented in AMPL and experiments carried out using the CPLEX solver on an
Intel Core i7-3770K CPU (3.5 GHz) with 8 virtual cores (4 cores with 2 threads
each), and 8 GB RAM. The parameters used for our experiments are shown in Table
\ref{tab:params}, and are based on the LuMaMi massive MIMO testbed
\cite{malkowsky2017world}, which has 100 antenna elements and uses a coherence
block structure similar to that used in LTE, that is, 12 OFDM subcarriers and 7
symbols per coherence block. In practice, a coherence block can
be larger, depending on the environment and device mobility, however these
parameters ensure coherence across time and frequency during the block for
practical scenarios of interest. The first symbol is used for pilots, resulting in 12
available pilots in each coherence block. We set the pilot length $S$
to 1, giving the minimum channel estimation quality and thus the most
challenging case for device scheduling.

In general, the large scale channel effects $\beta(k)$ may include both path
loss and large scale fading caused by, for example, shadowing from objects. (We
assume that channel hardening has eliminated small scale effects on the channel
\cite{gunnarsson2018channel}.) However, for this study we derive the values of
$\beta(k)$ only from path loss based on the distance of each user from the base
station. Since our optimization models rely only on the actual value of
$\beta(k)$, placing nodes at a greater distance from the base station, such that
they have greater path loss, is entirely equivalent to nodes placed closer to
the base station but subject to shadowing or other large scale effects that
reduce the channel gain. The path loss may be modeled as a function of the
distance $r(k)$ of each device $k \in \mathcal K$ to the base station, $\beta(k)
= \left(\frac{r(k)}{R}\right)^{-\alpha}$ for some path loss exponent $\alpha$
and reference distance $R$. Theoretically, the choice of reference distance is
immaterial. However, large differences in magnitude between the different
parameters can lead to floating point calculation errors while solving the
optimization problems, so we choose a reference distance that keeps the values
used in our scenarios within reasonable ranges.

For simplicity, we model the location of the base station as a single point,
that is, each end device has the same distance to all antenna elements. The SNR
and path loss exponent used are based on typical values for outdoor,
non line-of-sight transmission. For these experiments, we gave all devices the
same SINR threshold, and so in the following we will without loss of generality
refer to the threshold as simply $\mu$.
\begin{table}
    \begin{centering}
	\begin{tabular}{|c|c|}
	    \hline
	    \textbf{Parameter} & \textbf{Value} \\
	    \hline
	    Uplink SNR $\hat \rho$ & 10 dB \\
	    \hline
	    Downlink SNR $\check \rho$ & 10 dB \\
	    \hline
	    Number of antennas $M$ & 100 \\
	    \hline
	    SINR threshold $\mu$ & 1.0 (0 dB) \\
	    \hline
	    \hline
	    Pilot length $S$ & 1 \\
	    \hline
	    Number of pilots $P$ & 12 \\
	    \hline
	    Path loss exponent $\alpha$ & 3.7 \\
	    \hline
	    Reference distance $R$ & 200 m \\
	    \hline
	    Number of users $K$ & 40 \\
	    \hline
	\end{tabular}
	\caption{Parameters used for experiments.}
	\label{tab:params}
    \end{centering}
\end{table}
\begin{table}
    \begin{centering}
	\begin{tabular}{|c|c|c|c|}
	    \hline
	    \textbf{Experiment} & \newlinecell{\textbf{Near}\\\textbf{distance}} &
	    \newlinecell{\textbf{Far}\\\textbf{distance}} & \newlinecell{\textbf{Other}\\\textbf{parameters}} \\
	    \hline
	    1 & 50 m & 200 m & \\
	    \hline
	    2 & 200 m & 400 m & \\
	    \hline
	    3 & 50 m & 100 m & \\
	    \hline
	    4 & 50 m & 100 m & \newlinecell{$\mu$: 1, 5\dots50; step 5,\\ $K$: 20} \\
	    \hline
	    5 & 50 m & 100 m & $K$: 4\dots40, step 4 \\
	    \hline
	    6 & 50 m & 500 m & $K$: 40 total: 8 near, 32 far \\
	    \hline
	\end{tabular}
	\caption{Experiment configurations.}
	\label{tab:experiments}
    \end{centering}
\end{table}

We conducted six different experiments, according to the configurations shown
in Table \ref{tab:experiments}. Parameters not mentioned in the table are as
shown in Table \ref{tab:params}. Experiments 1--3 explore different near and far
distances for the device groups, while Experiment 4 tests the effect of the SINR
threshold, and Experiment 5 varies the number of devices. Experiment 6 tests an
unbalanced scenario in which there are a small number of nodes with good channel
conditions (close to the base station), and a larger group of nodes with much
poorer channels (far from the base station).

We defined six different scenarios for the experiments, shown in Table
\ref{tab:scenarios}. In each scenario, the nodes are divided into two groups, a
near group and a far group, with $\frac{K}{2}$ nodes in each group, except for
Experiment 6, where there are $\frac{K}{5}$ nodes in the near group, and
$\frac{4K}{5}$ nodes in the far group. The distances of the near and far groups
from the base station are different for different experiments, but in each case
the near group is closer. There are two traffic demand levels, high demand (10
coherence blocks) and low demand (2 coherence blocks). The scenarios provide
different combinations of low and high uplink and downlink demands for nodes
close to and far from the base station. The distances of the devices are
important for power control, since close devices will have a stronger signal
than far devices.
\begin{table}
    \begin{centering}
	\begin{tabular}{|c|c|c|}
	    \hline
	    \multicolumn{3}{|l|}{\textbf{Scenario 1}} \\
	    \hline
	    \textbf{Group} & \textbf{Uplink demand $\hat h$} & \textbf{Downlink demand $\check h$} \\
	    \hline
	    Near & 10 & 10 \\
	    \hline
	    Far & 2 & 2 \\
	    \hline
	    \multicolumn{3}{|l|}{\textbf{Scenario 2}} \\
	    \hline
	    \textbf{Group} & \textbf{Uplink demand $\hat h$} & \textbf{Downlink demand $\check h$} \\
	    \hline
	    Near & 2 & 2 \\
	    \hline
	    Far & 10 & 10 \\
	    \hline
	    \multicolumn{3}{|l|}{\textbf{Scenario 3}} \\
	    \hline
	    \textbf{Group} & \textbf{Uplink demand $\hat h$} & \textbf{Downlink demand $\check h$} \\
	    \hline
	    Near & 2 & 10 \\
	    \hline
	    Far & 10 & 2 \\
	    \hline
	    \hline
	    \multicolumn{3}{|l|}{\textbf{Scenario 4}} \\
	    \hline
	    \textbf{Group} & \textbf{Uplink demand $\hat h$} & \textbf{Downlink demand $\check h$} \\
	    \hline
	    Near & 10 & 2 \\
	    \hline
	    Far & 2 & 10 \\
	    \hline
	    \multicolumn{3}{|l|}{\textbf{Scenario 5}} \\
	    \hline
	    \textbf{Group} & \textbf{Uplink demand $\hat h$} & \textbf{Downlink demand $\check h$} \\
	    \hline
	    Near & 10 & 2 \\
	    \hline
	    Far & 10 & 2 \\
	    \hline
	    \multicolumn{3}{|l|}{\textbf{Scenario 6}} \\
	    \hline
	    \textbf{Group} & \textbf{Uplink demand $\hat h$} & \textbf{Downlink demand $\check h$} \\
	    \hline
	    Near & 2 & 10 \\
	    \hline
	    Far & 2 & 10 \\
	    \hline
	\end{tabular}
	\caption{Scenarios for experiments.}
	\label{tab:scenarios}
    \end{centering}
\end{table}

We tested all six scenarios for each experiment configuration For all
experiments, we found the minimal frame size for each of the power control
schemes (optimal, fair, and static). We also tested optimal power control with
no uplink power control, which we will hereafter call the downlink power control
scheme. For downlink power control, all uplink power control coefficients were
set to 1.0 (0 dB). Both maximum ratio combining and zero forcing were evaluated.
Our experiments generated a large amount of data, although in many cases the
results were as would be expected and/or were similar across the different
scenarios and experiments. Because of this, in the following sections, we give
only a summary of key experimental results. The full results are available in
\cite{report}. In our results, the power values obtained are not the actual
energy used by the nodes, but rather represent sums of the power control
coefficients used in the coherence blocks in the frame. Total power is thus the
sum of all the power control coefficients (both uplink and downlink) for all
nodes active in each allocated block. Similarly, max node power is the sum of
the power coefficients in all blocks in which a node was active, for the node
with the highest such sum, that is, the total power for the node with the
highest total power. This gives an initial comparison of the energy efficiency of
the different power control schemes. For a more comprehensive analysis, an
energy model would need to be applied to the allocated transmissions of the
nodes and the base station.

\subsection{Results}

We use a number of different performance metrics to evaluate our experimental
results. For reference, these are summarized in Table \ref{tab:metrics}.

\begin{table}
    \begin{centering}
	\begin{tabularx}{\columnwidth}{|l|X|}
	    \hline
	    \textbf{Metric} & \textbf{Definition} \\
	    \hline
	    Frame size & The objective of our optimization functions: the number
	    of coherence blocks required to satisfy all end devices' traffic
	    demands. Smaller frame sizes yield higher throughput.
	    \\
	    \hline
	    Total power & The sum of the transmission power control coefficients
	    for all end devices, for all coherence blocks in the frame. \\
	    \hline
	    Max node power & The sum of the transmission power control
	    coefficients, for all coherence blocks in the frame, for the end
	    device
	    that had the highest such sum. \\
	    \hline 
	    Solution time & The time, in seconds, required to solve each
	    optimization problem. \\
	    \hline 
	    Total solution time & The sum of the solution times of all
	    optimization problems solved to reach the final solution. \\
	    \hline
	\end{tabularx}
	\caption{Definition of performance metrics used in the experimental
	results.}
	\label{tab:metrics}
    \end{centering}
\end{table}

\subsubsection{Experiments 1--3} \label{subsubsec:exp-1-3}
First, we will discuss the results from Experiments 1--3. All the results shown
in the figures in this section are from Experiment 1, however, the results for
Experiments 2 and 3 were similar and are omitted from further discussion.

The minimum frame sizes obtained were 21 coherence blocks for scenarios 1 and 2,
and 35 blocks for scenarios 3--6. In some isolated cases, most often when using
static power control, the minimum frame size was greater by one coherence block.
This can be caused by more restrictive power control schemes being unable to
accommodate the c-sets needed for the smaller frame size, although in one case
this even occurred for optimal power control.  This is a result of the different
c-sets generated when solving the pricing problems.
As discussed in Sections~\ref{sububsec: Phase 2} and \ref{subsec:main-problem}, the c-sets needed to optimally solve the main
problem may differ from those needed to optimally solve its linear relaxation.
In some cases, this can result in (slightly) suboptimal final solutions.

Based on the scenarios tested, smaller frame sizes result from situations where
the device groups are well separated in terms of demand, that is, where the near
and far nodes do not compete --- all high demand nodes are at the same distance
to the base station and thus have similar channels. Whether high demand nodes
compete on the uplink or downlink does not appear to affect the frame size in
these cases.

All power control schemes gave similar performance in terms of frame size,
including downlink power control. This means that we can achieve similar throughput
performance without uplink power control, if we schedule the nodes effectively.
This can reduce signaling overhead as well as simplify the implementation of
resource-constrained IoT end devices. Transmitting at full power on the uplink
will of course consume more energy, however this may be mitigated by efficient
scheduling resulting in longer sleep times between transmissions. A full
investigation of energy efficiency is however beyond the scope of this paper.
\begin{figure}
    \begin{subfigure}{\columnwidth}
    \begin{centering}
	\includegraphics[width=\columnwidth]{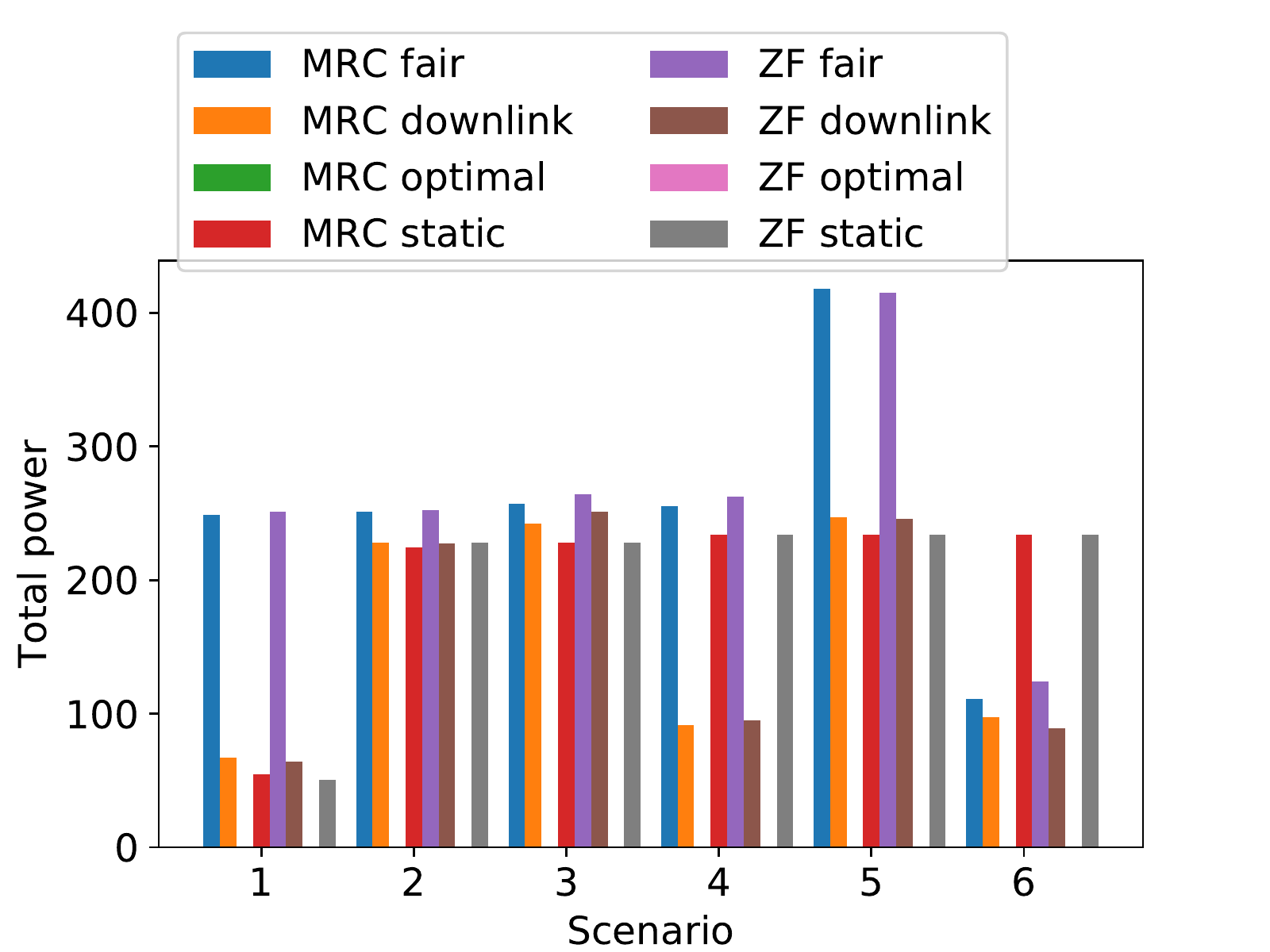}
	\caption{Total power, Experiment 1.}
	\label{fig:experiment1_total_power}
    \end{centering}
\end{subfigure}
\begin{subfigure}{\columnwidth}
    \begin{centering}
	\includegraphics[width=\columnwidth]{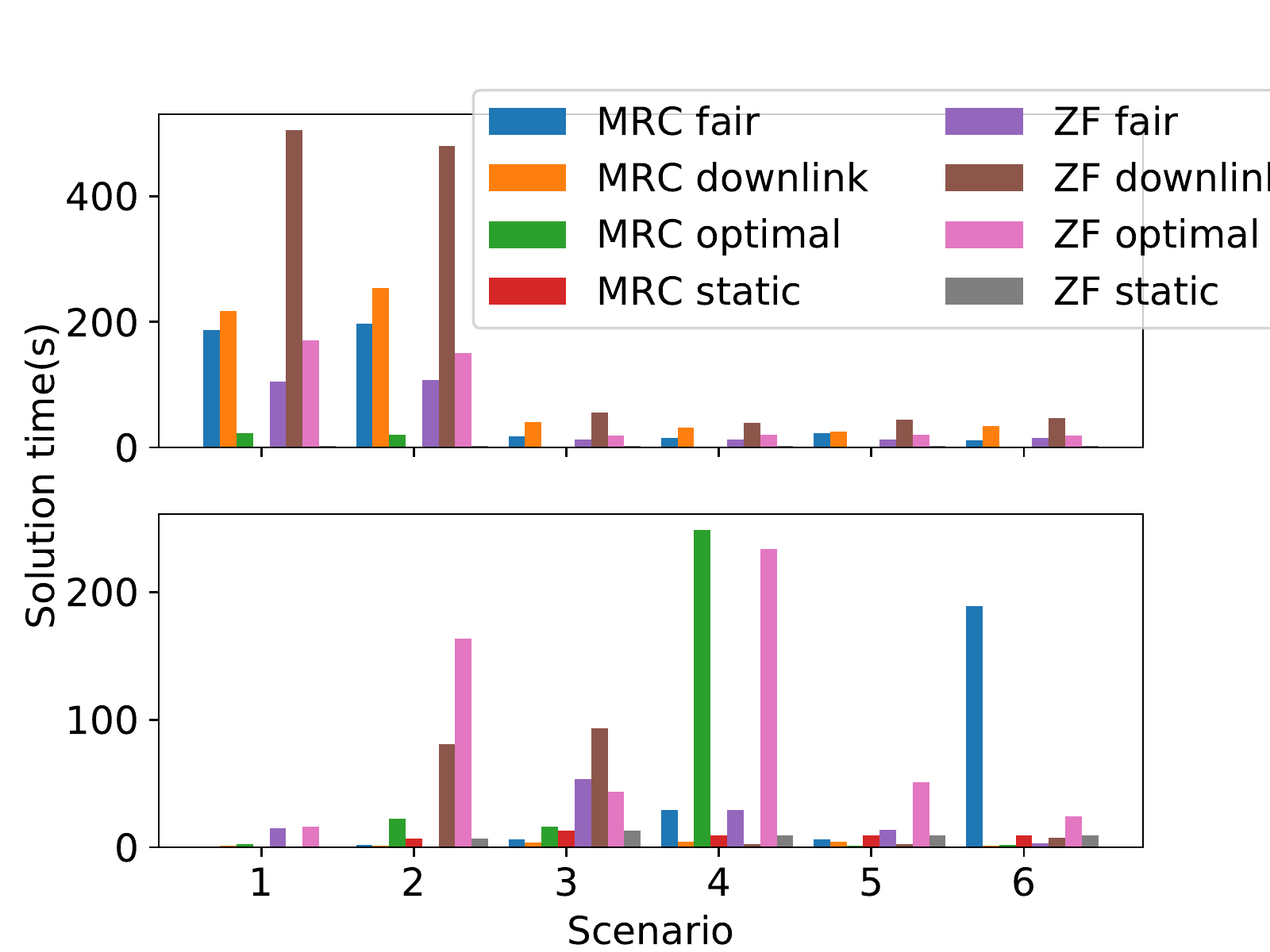}
	\caption{Solution time for the pricing problem (top) and the main problem
	(bottom), Experiment 1.}
	\label{fig:experiment1_solution_time}
    \end{centering}
\end{subfigure}
\caption{Experiment 1 results.}
\end{figure}

Nonetheless, the total power, shown in Figure \ref{fig:experiment1_total_power},
gives an indication of the comparative energy usage for the different
configurations tested. As would be expected, the closer the devices are to the
base station, the lower the power needed. Performing full optimal power control
can result in significant energy savings; note that the power values shown in
the figures for optimal power control are not zero, but rather very small
relative to the other values. Fair power control gives the worst energy
efficiency, even higher than static power control. This is because this power
control scheme was designed to give the maximum fair SINR to all devices. Static
power control, since it is fair power control performed over all devices, not just
those in the current c-set, can only result in a lower or equal SINR, and uses
less power for nodes with good channel quality. Both of these schemes in effect
over-estimate the transmission power needed, as they are not designed with
limited traffic demands in mind, but rather for saturated scenarios where all
devices seek to maximize their throughput. The maximum node power results, not
shown here, followed a similar pattern.

The number of pricing problem iterations in the CG algorithm required to generate all needed c-sets
was less than 100 in most cases, with notable exceptions being fair, downlink,
and optimal power control in scenarios 1 and 2, where more than 500 iterations
were needed. Figure \ref{fig:experiment1_solution_time} shows the total solution
time, broken down into the time needed for (all iterations of) the pricing
problem, and the time needed for the main problem, i.e., the final MP/IP($\mathcal C^*$). As
would be expected, static power control takes the least amount of time, as here
the power control coefficients are not decision variables to be optimized.
Optimal power control also performs quite well, in most cases solving faster
than fair power control.  This is because for optimal power control, the power
control constraints are linear, whereas fair power control requires additional
integer variables.  Scenarios 1 and 2, which had the smallest frames, required a
much higher number of pricing problem iterations for all power control schemes
except static power control.

In most cases, the main problem constitutes a large proportion of the solution
time, but in a few cases the pricing problem instead takes significant time.
However, the objective becomes close to its final, optimal, value after
relatively few iterations: fewer than 20 iterations for the frame size to drop
below 50 from initial values of between 200 and 400. This indicates that
a good, albeit suboptimal, list of c-sets can be achieved by only performing a small
number of iterations, which would dramatically reduce the time spent solving the
pricing problem.

\subsubsection{Experiment 4}
In Experiment 4, we varied the SINR threshold $\mu$. This can equivalently be
interpreted as worsening the channel quality of all the nodes, for example by
moving them further away from the base station. This allows us to investigate
how the frame size, power, and solution times are affected by solving the
optimization problem under more challenging channel conditions. The results
presented are from scenario 1, however the overall behavior observed in the
other scenarios as the SINR threshold was increased was similar.

\begin{figure}
    \begin{subfigure}{\columnwidth}
    \begin{centering}
	\includegraphics[width=\columnwidth]{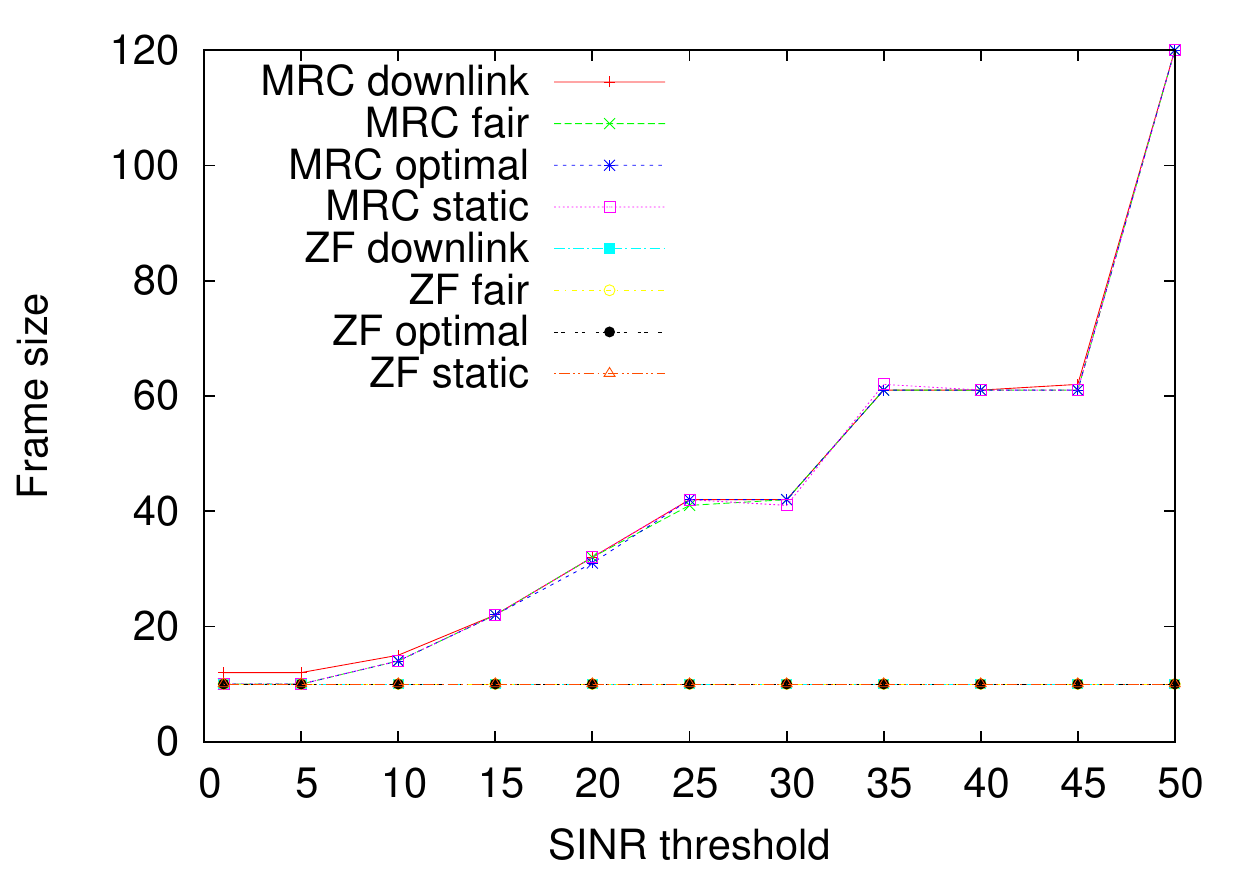}
	\caption{Frame size, Experiment 4, scenario 1.}
	\label{fig:experiment4_frame_size0}
    \end{centering}
\end{subfigure}
\begin{subfigure}{\columnwidth}
    \begin{centering}
	\includegraphics[width=\columnwidth]{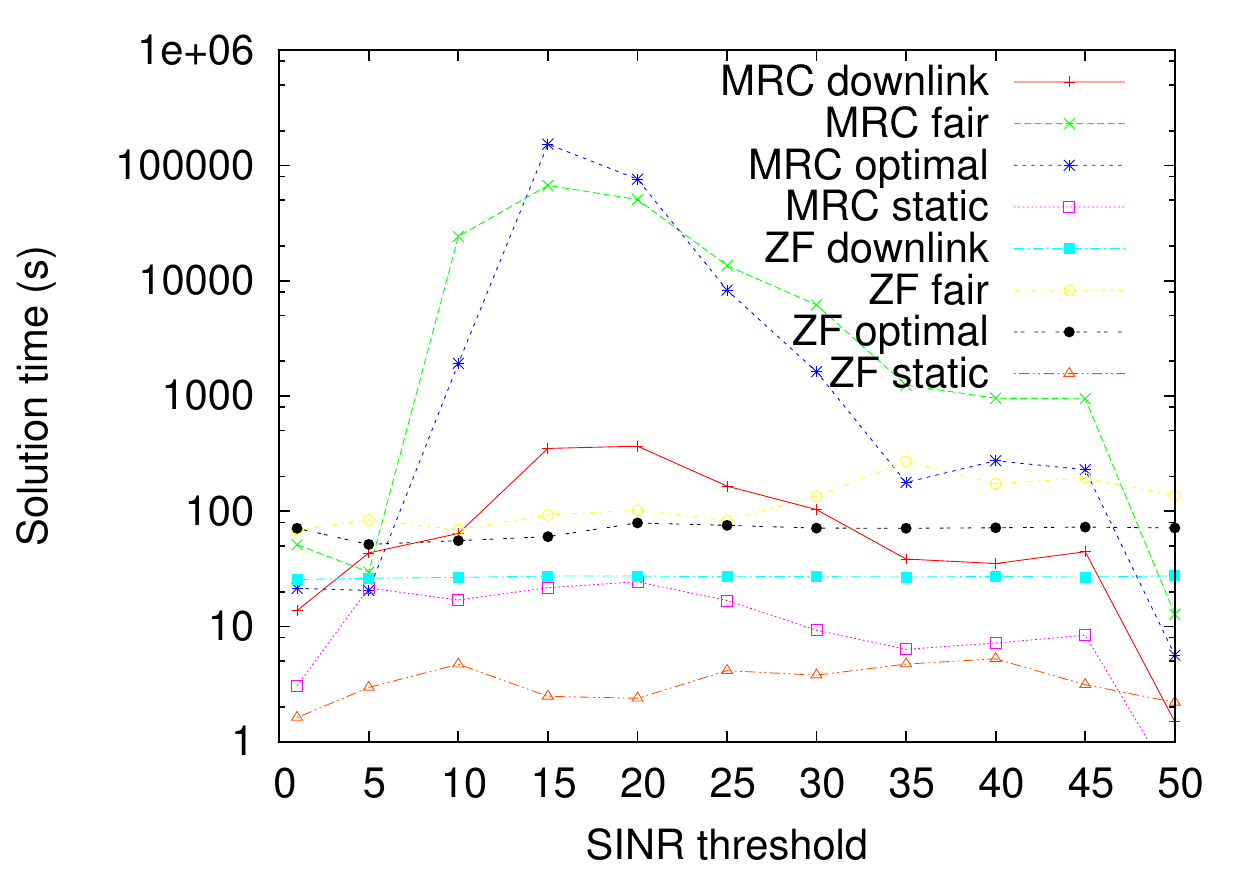}
	\caption{Total solution time, Experiment 4, scenario 1.}
	\label{fig:experiment4_solution_time0}
    \end{centering}
\end{subfigure}
\caption{Experiment 4 results.}
\end{figure}

Figure \ref{fig:experiment4_frame_size0} shows the frame size vs. SINR for all
power control and precoding schemes tested. From the figure, we can see that the
power control scheme makes little difference to the frame size over different
SINR threshold values, however the behavior for the two precoding schemes is
very different. With maximum ratio combining, the frame size increases
monotonically with increasing SINR threshold. We can observe regions where the
frame size is held steady, before jumping up to its next value; this is because
the SINR threshold reaches a critical point where one or more devices are no
longer able to be accommodated in the same c-set, resulting in a different
scheduling configuration with a larger frame.

However, zero forcing is able to maintain a steady frame size across all SINR
threshold values, and even for higher threshold values tested in experiments not
reported here. This is because the effectiveness of zero forcing, unlike MRC,
does not depend on the relative quality of the devices' channels, but rather on
the accuracy of the channel estimation, which is not affected by increasing the
SINR threshold. Zero forcing is thus able to maintain simultaneous transmission
using the c-sets needed for the optimum frame size, essentially up until the point
where the SINR threshold is so high that transmission is not possible to the
devices at all. In terms of total and max node power,
we again observed the same dichotomy between the two precoding schemes, with zero
forcing maintaining steady performance across all SINR threshold values tested,
while for MRC the power increases with increasing SINR threshold.

Figure \ref{fig:experiment4_solution_time0} shows the total solution time
as the SINR threshold is varied. Here we see a dramatic spike
in solution times for intermediate SINR threshold values, especially for MRC
with fair and optimal power control, but even with the other power
control schemes. For ZF, the solution times are more consistent, but do vary
with the power control scheme. The increase in solution time for MRC is likely
due to the larger number of possible solutions in this region that need to be
eliminated by the solver during branch-and-bound. In the intermediate SINR
threshold region, there are many candidate c-sets where some or all nodes would achieve
SINR close to the threshold value for some or all power coefficient values.
Meanwhile, at high SINR thresholds, there are few c-sets that can satisfy the
SINR constraints; in fact, at very high SINR thresholds, only singleton c-sets
are possible, where only one device is served at a time. In the low SINR threshold
region, most or all nodes are easily able to be accommodated in the same c-set.
Increased solution times in the intermediate SINR region occurred for
both the main and pricing problems. However, these longer solution times were
not due to an increased number of iterations of the pricing problem. Rather, the
individual iterations took longer to solve.

These results indicate that, particularly for MRC, optimal solutions may be most
suitable for cases where devices have either quite high or quite poor channel
quality, whereas for intermediate cases, other approaches may be more suitable.
Such approaches could include the variations on our solution algorithm discussed
in Section~\ref{subsec:main-problem}, or other approximation algorithms. The
development of such algorithms will be the subject of our future work.

\subsubsection{Experiment 5}
\begin{figure}[h]
    \begin{centering}
	\includegraphics[width=\columnwidth]{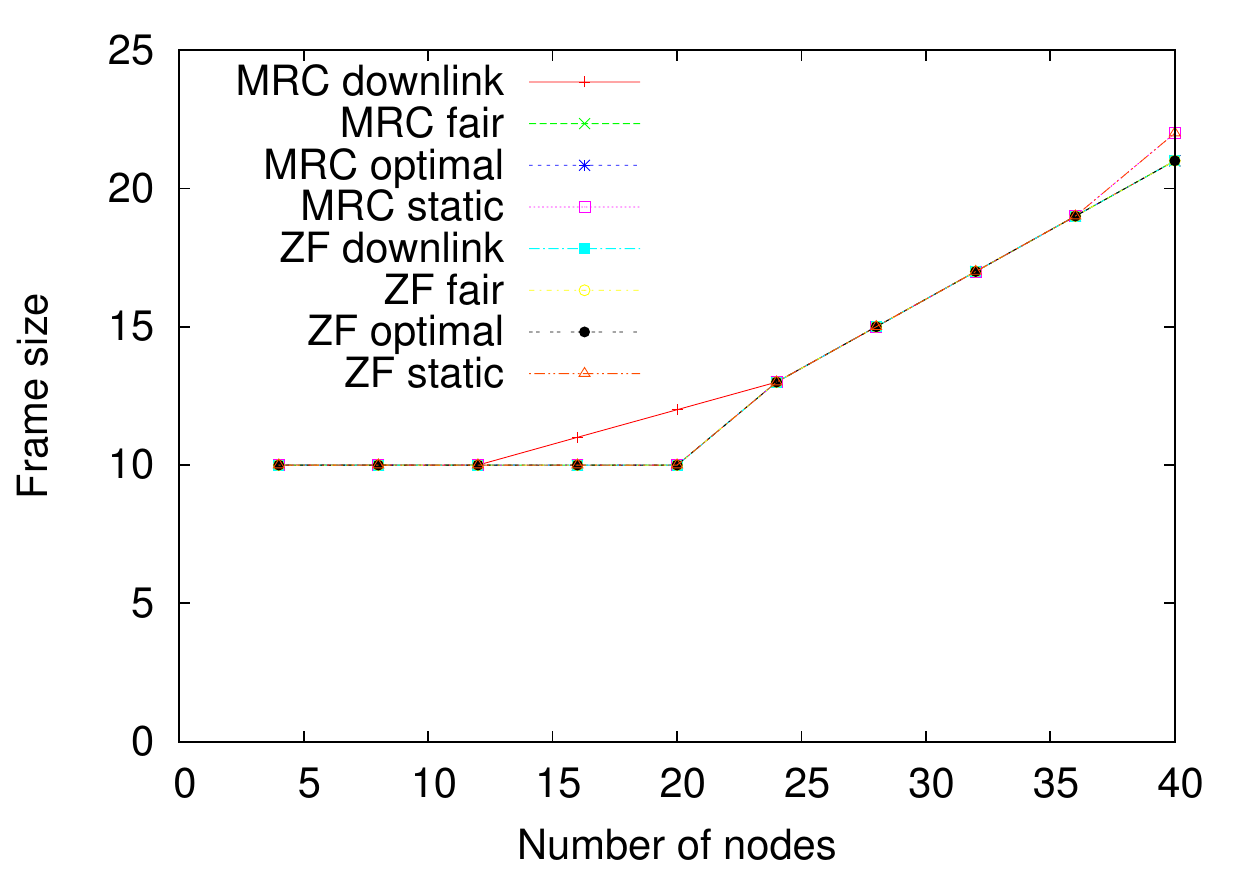}
	\caption{Frame size, Experiment 5, scenario 1.}
	\label{fig:experiment5_frame_size0}
    \end{centering}
\end{figure}

In Experiment 5, we varied the number of devices. Again, the results
presented here concern scenario 1, but similar behavior was observed for the
other scenarios. Figure \ref{fig:experiment5_frame_size0} shows the frame size
as the number of devices increases. Below a certain point, more devices can be added
without increasing the minimum frame size, however after that the frame size
increases with the number of devices. This is unsurprising, since with a fixed SINR
threshold the viable c-sets and thus frame size are largely
determined by the number of available pilots.

In terms of the power used, the total power increased steadily with the number
of devices, while the max node power did not show any clear trend in relation to
the number of devices. The total solution time increased exponentially with the
number of devices, and this was the case for the time for both the pricing problem
and the main problem. This behavior is typical of this kind of optimization
problem. The number of pricing problem iterations also increased with the number
of devices, however here there was more variation.

\subsubsection{Experiment 6}
\begin{figure}
    \begin{subfigure}{\columnwidth}
    \begin{centering}
	\includegraphics[width=\columnwidth]{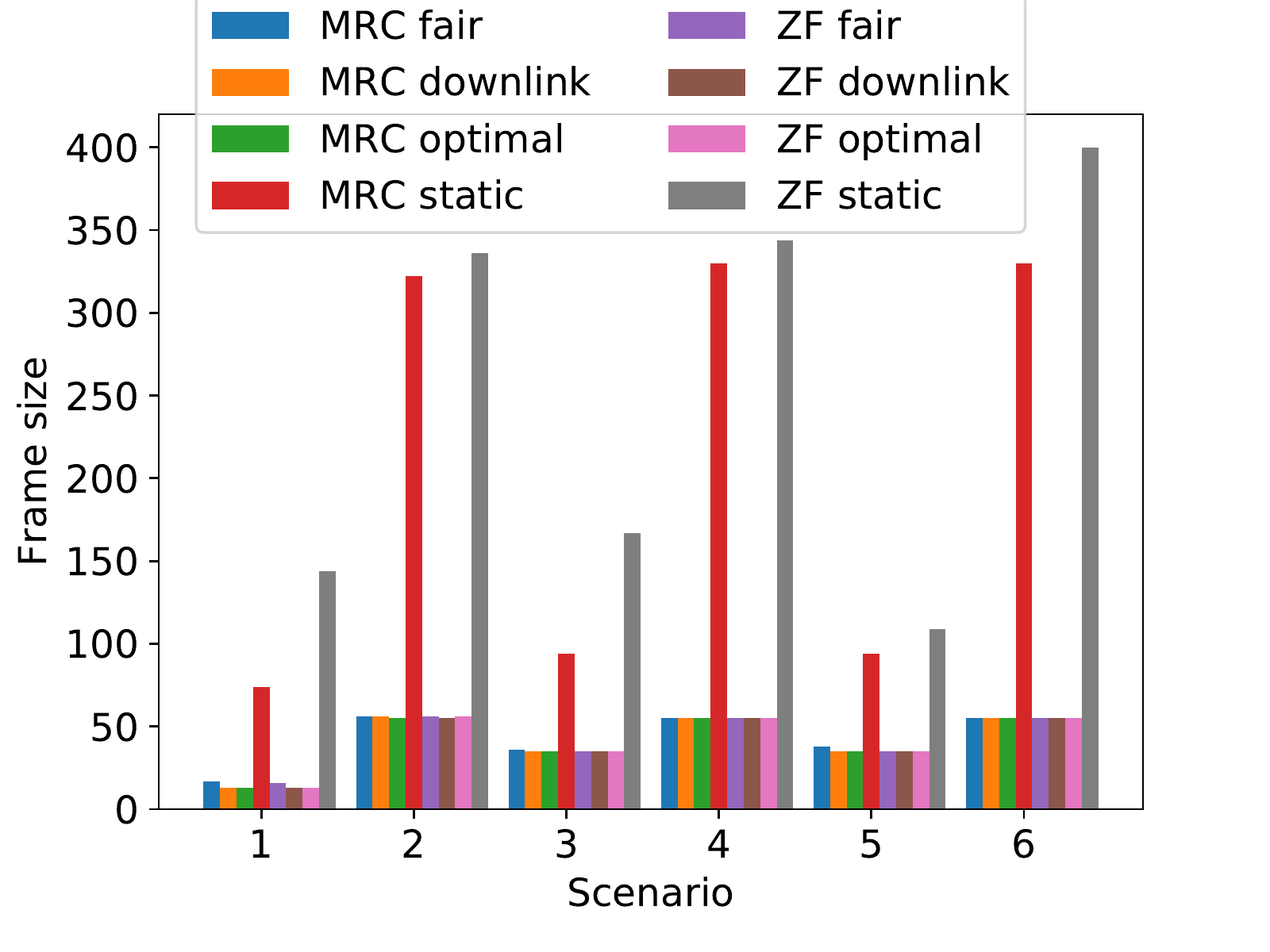}
	\caption{Frame size}
	\label{fig:experiment6_frame_size}
    \end{centering}
\end{subfigure}
\begin{subfigure}{\columnwidth}
    \begin{centering}
	\includegraphics[width=\columnwidth]{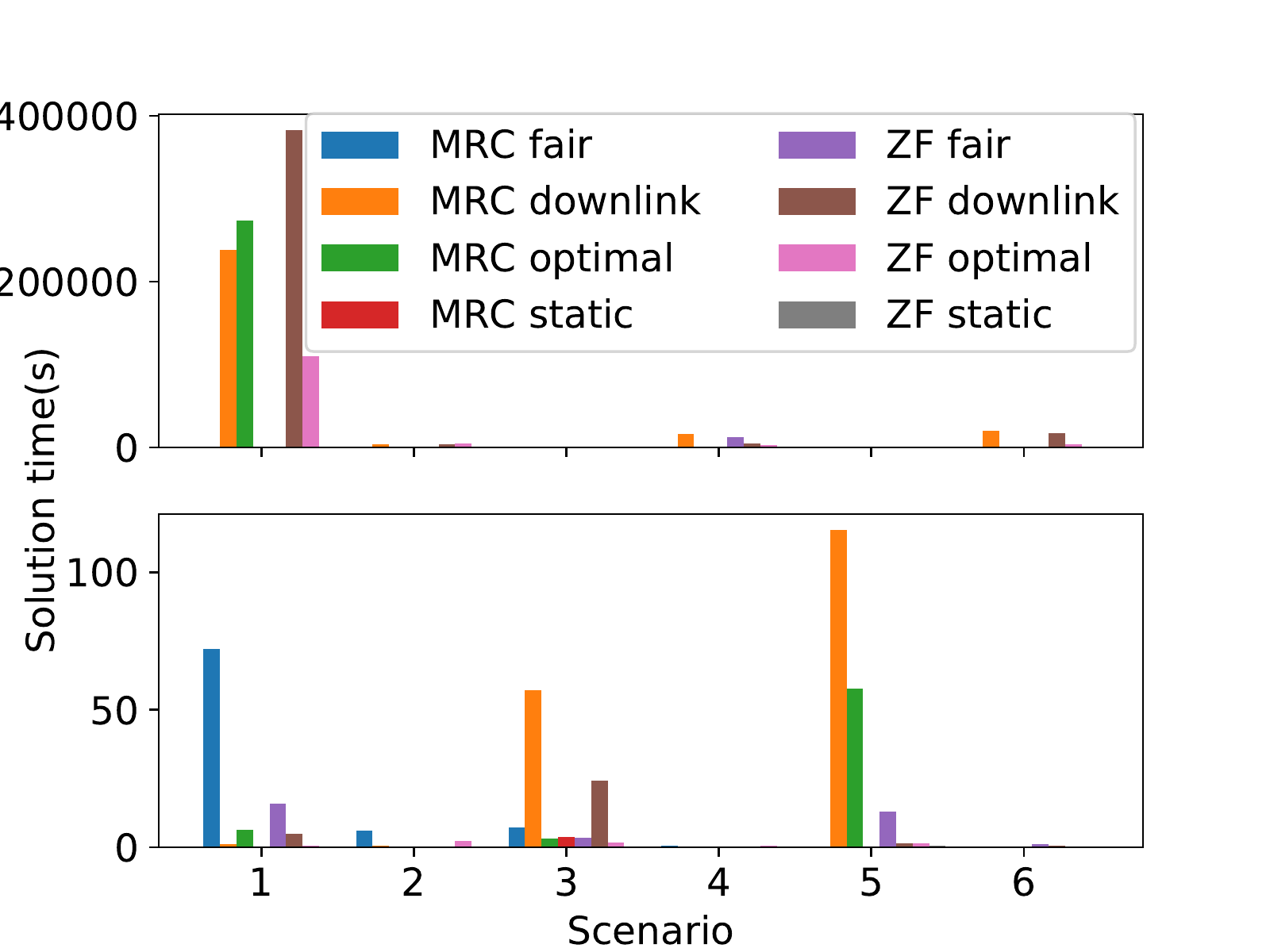}
	\caption{Solution time for the pricing problem (top) and main problem
	(bottom)}
	\label{fig:experiment6_solution_time}
    \end{centering}
\end{subfigure}
\caption{Experiment 6 results.}
\end{figure}

In Experiment 6, since the near and far groups are unbalanced, with four times
as many nodes in the near group as in the far group, the frame sizes (Figure
\ref{fig:experiment6_frame_size} were more varied between the different scenarios
than in the other experiments. Here, the frame size was primarily determined by
the total downlink demand, so that in scenarios where the larger (far) group had
higher downlink demands, the frame size was also larger. This is because in this
experiment, the far group had a very poor channel, making it more difficult for
the base station to share its power across the different simultaneous devices in
such a way as to achieve an acceptable SINR for all devices.

This is particularly evident when looking at the frame sizes when using static
power control. With this power control scheme, the base station effectively
allocates some of its transmission power to the nearby nodes, even when they are
not scheduled to receive any data in a given coherence block. The remaining
power allocated to the far nodes is then not enough to achieve a good SINR,
preventing those nodes from being scheduled simultaneously, and resulting in
much larger frame sizes. With power control optimization, however, the base
station tailors its power allocation to the nodes that are actually active in
each block, and is thus able to provide more power (and therefore a higher SINR)
to the far nodes when they are scheduled on their own, without any close nodes
active. The long frame sizes in these cases also resulted in higher total and
max node power. For such a scenario, then, where the majority of nodes have poor
channels, and especially when there are large downlink demands, transmission
power control optimization becomes very important to ensure the efficient use of
resources.

The solution times for Experiment 6 (Figure \ref{fig:experiment6_solution_time})
were mostly short, except for Scenario 1. In this case, the near group had
higher demands than the far group. As discussed above, scheduling even one node
from the near group at the same time as nodes from the far group significantly
affects the transmission power allocation. This results in a more difficult
scheduling problem when the majority of the traffic to be scheduled belongs to
the near group.

\subsubsection{Heuristic Methods}\label{sec:heuristic_results}
Table~\ref{tab:heuristics} shows the performance of the different phases of our
solution approach for Experiment 6, Scenario 1, as well as the performance when
solving the main problem on a selected family of c-sets $\mathcal C'$ consisting
of only those used in the optimal solution to the linear relaxation (rightmost
column). From the table, we can see that the lower bound provided by the linear
relaxation is very good, deviating by only two coherence blocks at most from the
final IP solution of MP/IP($\mathcal C^*$). Our proposed heuristic, using only c-sets in $\mathcal C'$,
also achieved objective values very close to those using the entire generated
family $\mathcal C^*$, while substantially reducing the time required to solve the
main problem. These times are short, both for the heuristic and the full MP/IP($\mathcal C^*$),
thanks to the high quality lower bound produced by the linear relaxation.
However, as seen in the results from Experiment 5, the times will grow
exponentially as the size of the network grows. As the solution times increase,
a timeout can be used to attempt solution of the main problem for family $\mathcal C^*$,
and fall back to using only selected c-sets, i.e., family $\mathcal C'$, if the main problem cannot be solved
in a reasonable time.
\begin{table*}
    \begin{centering}
	\begin{tabular}{|c|c|c|c|c|c|c|c|c|c|}
	    \hline
	    \textbf{Configuration} & \newlinecell{\textbf{Time} \\ \textbf{MP/LR}} & $|\mathcal C^*|$
	    & \newlinecell{\textbf{Objective} \\ \textbf{MP/LR}} & \newlinecell{\textbf{Time} \\ \textbf{MP/IP($\mathcal C^*$)}} & \newlinecell{\textbf{Objective} \\ \textbf{MP/IP($\mathcal C^*$)}} &
	    \newlinecell{\textbf{Time}\\ \textbf{MP/IP($\mathcal C'$)}} & $|\mathcal C'|$ &
	    \newlinecell{\textbf{Objective}\\ \textbf{MP/IP($\mathcal C'$)}} \\
	    \hline
	    MRC downlink & 0.045 & 359 & 15.333 & 14.016 & 17 & 0.051 & 45 & 17
	     \\
	    \hline
	    MRC fair & 0.033 & 518 & 12.245 & 0.341 & 13 & 0.074 & 67 & 14 \\
	    \hline
	    MRC optimal & 0.053 & 572 & 12.235 &1.347 & 13 & 0.123 & 67 & 14 \\
	    \hline
	    MRC static & 0.015 & 46 & 74.0 & 0.015 & 74 & 0.016 & 33 & 74 \\
	    \hline
	    ZF downlink & 0.021 & 247 & 14.0 & 2.790 & 16 & 0.143 & 70 & 16 \\
	    \hline
	    ZF fair & 0.053 & 526 & 12.235 & 1.074 & 13 & 0.069 & 61 & 14 \\
	    \hline
	    ZF optimal & 0.041 & 517 & 12.235 & 0.177 & 13 & 0.068 & 62 & 14 \\
	    \hline
	    ZF static & 0.007 & 43 & 144.0 & 0.015 & 144 & 0.018 & 40 & 144 \\
	    \hline
	\end{tabular}
    \caption{Results for Experiment 6, Scenario 1, using all generated c-sets ($\mathcal C^*$)or
    only those required for the optimal linear solution ($\mathcal C'$). All times are in seconds.}
    \label{tab:heuristics}
\end{centering}
\end{table*}

\subsection{Discussion}

Somewhat surprisingly, most of the experiments we conducted did not give any
substantial differences in frame size between either the different power control
schemes, or the different precoding schemes. However, the other performance
measures investigated, namely power and solution time, do vary greatly, which
indicates there are benefits to choosing one scheme over another depending on
the specific network configuration. In the case of a large group of nodes with a
poor channel, optimizing transmission power did show substantial benefits in
frame size, with static power control performing much worse than the other power
control schemes. This is a fairly typical, realistic scenario, especially in
situations such as urban environments where there is significant variation in
channel gains resulting in a long tail distribution of channel gain among the
devices. Our results here show a substantial improvement in throughput by
jointly optimizing power control along with device scheduling.

Although a full energy model is needed to obtain concrete energy values, using
optimal power control provided clear benefits in terms of energy savings. This
is true for both the total power of the whole network, as well as the maximum
individual power for any device node. Zero forcing with optimal power control
consistently used the least power, two orders of magnitude lower than the others
in most cases for Experiment 4, and three orders of magnitude in the case of
Experiment 5. However, our experiments are insufficient to provide comprehensive
guidance on which precoding scheme to use, since some important aspects of their
performance, as well as intermediate schemes such as minimum mean-square error
and regularized zero-forcing, are not considered here. For example, some
research has shown that MRC outperforms ZF for multi-cell systems, depending on
the number of devices and antennas, and the pilot reuse factor
\cite{bjornson2016massiveb}.

The fair power control schemes widely adopted in the literature on massive MIMO
(see Section \ref{sec:related_work}) are targeted towards use cases with
homogeneous devices and traffic, where the primary goal is high throughput. Our
results here however show clear benefits in tailoring power control for
heterogeneous devices and traffic demands, especially where energy efficiency is
of concern, as is usually the case for IoT scenarios, or where there are large
discrepancies in channel qualities between devices. IoT devices are also often
resource constrained, and our results show that power control on the uplink may
be avoided by employing efficient scheduling without sacrificing throughput.
This interaction between scheduling and power control is not taken into account
in previous schemes, but rather the uplink and downlink are treated similarly.

\section{Future Work}\label{sec:future_work}
There are many possible extensions to this work, both to further develop and
validate our models, as well as to apply them to other problems and application
scenarios. Here we have tested systematically constructed scenarios intended to
illuminate how the performance of both the underlying massive MIMO system and
our optimization formulations changes with different system parameters. An
important step for future work is then to also test our formulations with
real network traces and/or randomly generated network scenarios. The channel models could
also be made more realistic, as those used here are relatively simple, with only
path loss considered in our experiments, and only large scale effects taken into
account in our model. In future work, channel models for multi-cell systems
could also be used. This would significantly increase the complexity of the
models, but would allow for performance of larger massive MIMO systems to be
studied, as well as phenomena such as inter-cell cooperation and pilot
contamination.

Energy usage is a critical performance measure for many IoT systems, and here we
have not optimized for energy efficiency, but rather for overall system
throughput. However, our model can also be applied to different objective
functions, including minimal energy usage, fair energy usage, or delay
minimization, which may be important for mission critical systems or tactile
internet applications. In order to accommodate these new objectives, new
versions of the main problem need to be formulated that model these performance
measures, and from there the dual problems and thus pricing problem objectives
can be derived. The pricing problem constraints we have formulated here, and
which constitute the most difficult part of the formulations, will however still
apply.

In our experiments we have investigated the solution times to find the optimal
frame size. In some application
scenarios, it would be feasible to use the optimal solutions, or a modified
version of our solution algorithm where the time to solve is improved by either
reducing the number of iterations, applying time limits, or using heuristic
methods as discussed in Section \ref{subsec:main-problem}. This is especially
true in the case of periodic traffic, where the optimization problem need only
be solved once, and the solution can then be used for a long time. However, for
cases where it is not feasible to apply our optimization formulations, there is
a need to develop new, more efficient approximation algorithms.

\section{Conclusion}\label{sec:conclusion}

In this paper we have developed a new model adapting the concept of compatible
sets to massive MIMO systems, which allows for the efficient solution of a
variety of types of optimization problems relating to network
performance. We have applied our model to the case of joint device
scheduling and power control for maximum throughput, considering two different
precoding schemes and three power control schemes. Our results show substantial
benefits in terms of energy usage when treating the power control coefficients
as optimization variables, and large gains in throughput in jointly optimizing
power control and device scheduling in scenarios with a large spread in channel
qualities between devices. On the other hand, much simpler power control can
also be applied without loss of throughput. In particular, the same throughput
can be achieved without performing any power control on the uplink at
all, which may reduce the complexity needed in resource-constrained
IoT end devices, as well as reduce the signaling overhead
between the base station and the devices.

With the advent of 5G driving the adoption of massive MIMO, there is a need for
new modeling to analyze the performance of these systems, as well as provide
practical solutions for implementation, for the broad range of different
application scenarios and performance goals encompassed under the umbrella of
the 5G requirements. In this work, we have targeted the case of a large number
of devices with heterogeneous demands, and our models provide a flexible and
general method for network optimization in such scenarios.

\section*{Acknowledgements}
The work of Emma Fitzgerald and Fredrik Tufvesson was supported by the strategic research area
ELLIIT. The work of Micha{\l} Pi\'{o}ro (and partly of Emma Fitzgerald) was supported by the National
Science Centre, Poland, under the grant no. 2017/25/B/ST7/02313: ``Packet
routing and transmission scheduling optimization in multi-hop wireless networks
with multicast traffic''.

\appendix \label{appendix}

In order to remove the variable multiplications in formulations \eqref{PP-MRC},
\eqref{PP-ZF}, \eqref{MRC_fair}, and \eqref{ZF_fair}, we need to introduce the
following auxiliary variables and constraints. In formulation \eqref{PP-MRC},
we introduce auxiliary variables $\hat x_k,\,\check x_k \in \mathbb R_+,\, k \in
\mathcal K$, for the uplink and downlink respectively.  Constraints
\eqref{uplink-SINR-MRC} are then replaced with the following constraints:
\begin{subequations}
    \begin{align}
	& \Delta \left(1 - \hat{u}_k \right) + M \hat{\rho} \gamma(k)
	\hat{\eta}_k \geq \nonumber \\
	& \qquad\qquad\qquad \mu(k) \left( 1 + \hat{\rho} \sum_{k' \in \mathcal
	K} \beta(k') \hat{x}_{k'} \right), & k \in \mathcal K \\
	& \mathrlap{\hat x_k \leq \hat u_k; \,
	     \hat x_k \leq \hat \eta_k; \,
	 \hat x_k \geq \hat \eta_k + \hat u_k - 1 + \Delta (\hat u_k - 1),}
	\nonumber \\
	& & k \in \mathcal K. \label{eq:temp-app}
    \end{align}
\end{subequations}
Variable $\hat x_k$ thus represents the product $\hat u_k \hat\eta_k$. For a valid power
control scheme, $\hat \eta_k$ should always be at most 1, however, in fair power
control, this does not necessarily hold in cases where $u_k = 0$:
a node can be assigned a power control coefficient greater than 1 when the node
is not included in the c-set. This is because the coefficients
$\hat \eta_k,\,\check \eta_k$ are defined for all nodes (as all nodes can
potentially be included in a c-set), but their values are calculated relative to
the node in the c-set with the worst channel, that is, the lowest value of
$\gamma(k)$ such that $\hat u_k = 1$. This means that nodes which have lower
values of $\gamma(k)$ than any of the nodes in the c-set will have $\hat \eta_k
> 1$.

While this would have no affect on practical power control --- nodes with $\hat
\eta_k > 1$ do not transmit --- it would cause some of the constraints in
\eqref{eq:temp-app} to be infeasible (no acceptable value can be found for
$\hat x_k$) without the final term in \eqref{eq:temp-app}, which
simply cancels the lower bound on $\hat x_k$ when $\hat u_k = 0$. In such cases
$\hat x_k$ will be forced to zero.

Constraints \eqref{downlink-SINR-MRC} are replaced with
\begin{subequations}
    \begin{align}
	& \Delta \left(1 - \check{u}_k \right) + M \check{\rho} \gamma(k)
	\check{\eta}_k \geq \nonumber \\
	& \qquad\qquad \mu(k) \left( 1 + \check{\rho} \beta(k)
    \sum_{k' \in \mathcal K} \check{x}_{k'} \right), & k \in \mathcal K \\
	    & \check x_k \leq \check u_k; \, \check x_k \leq \check \eta_k; \,
	\check x_k \geq \check \eta_k + \check u_k - 1. & k \in \mathcal K.
    \end{align}
\end{subequations}
Here, the extra term is not required as $\check \eta_k$ will always lie in the
interval $[0,1]$.
\balance

In the case of zero-forcing, we replace constraints \eqref{uplink-SINR-ZF} and
\eqref{L0-ZF} with the following:
\begin{subequations}
    \begin{align}
	& \Delta \left(1 - \hat{u}_k\right) + \hat{\rho} \gamma(k) \left(
    M\hat{\eta}_k - \hat L_k \right)  \geq \nonumber \\
    & \mathrlap{\quad \mu(k) \left( 1 + \hat{\rho} \sum_{k' \in \mathcal K} \left( \beta(k') -
	    \gamma(k') \right) \hat{x}_{k'k'} \right),} & \qquad\quad k \in \mathcal K \\
	& \hat L_k = \sum_{k' \in \mathcal K} \hat x_{kk'}, & k \in \mathcal K \\
	& \mathrlap{\hat x_{kk'} \leq \hat u_k; \, \hat x_{kk'} \leq \hat \eta_{k'};
	\hat x_{kk'} \geq \hat \eta_{k'} + \hat u_k - 1 + \Delta (\hat u_k - 1),}
	\nonumber \\
	& &  k, k' \in \mathcal K. \label{zf_fair_aux3}
    \end{align}
\end{subequations}
Again, we need to consider the case where $\hat \eta_k > 1$ for nodes with $\hat
u_k = 0$, which is handled as for the MRC case with the final term in 
\eqref{zf_fair_aux3}. For the downlink, we replace constraints
\eqref{downlink-SINR-ZF} and \eqref{L1-ZF} with the following:
\begin{subequations}
    \begin{align}
	& \Delta \left(1 - \check{u}_k \right) + \check{\rho}
	\gamma(k)\left( M\check{\eta}_k - \check L_k \right) \geq \nonumber \\
	& \mathrlap{\quad \mu(k) \left( 1 + \check{\rho} \left( \beta(k) - \gamma(k) \right)
	\sum_{k' \in \mathcal K} \check{x}_{k'k'} \right),} & k \in \mathcal K  \\
	& \check L_k = \sum_{k' \in \mathcal K} \check x_{kk'}, & \quad \quad \; \; \, k \in \mathcal K \\
	& \mathrlap{\check x_{kk'} \leq \check u_k; \, \check x_{kk'} \leq \check
	\eta_{k'}; \,
    \check x_{kk'} \geq \check \eta_{k'} + \check u_k - 1,} \nonumber \\
    & & k, k' \in \mathcal K.
    \end{align}
\end{subequations}

Variable multiplications also occur in the formulations for fair power control.
We thus introduce auxiliary uplink variables $\hat y_k \in \mathbb B,\,k \in \mathcal
K$, and, in the case of MRC, auxiliary downlink variables $\check y_{kk'} \in
\mathbb R_+,\, k, k' \in \mathcal K$. Note that the uplink variables are binary,
while the downlink variables are real-valued. In the case of zero-forcing, we do
not need auxiliary variables for the downlink, since we can use variables
$x_{kk'}$ that were already required for the main pricing problem.

We then replace constraints \eqref{z-1-MRC} and \eqref{z-2-MRC} with the following:
\begin{subequations}
    \begin{align}
	& \sum_{k \in \mathcal K} \hat{y}_k \leq 1 \\
	& \sum_{k' \in \mathcal K} \hat{y}_{k'} \geq \hat{u}_k, & k \in
	\mathcal K \\
	& \hat y_k \leq \hat u_k; \, \hat y_k \leq z_k; \,
	\hat y_k \geq \hat u_k + z_k - 1. & k \in \mathcal K.
    \end{align}
\end{subequations}

For MRC, we replace constraints \eqref{downlink-power-MRC} with
\begin{subequations}
    \begin{align}
	& \mathrlap{\check{\rho} \gamma(k) \left( \frac{1}{\check{\rho}} \sum_{k' \in \mathcal
	    K} \frac{\check{y}_{kk'}}{\gamma(k')} + \sum_{k' \in \mathcal K}
	\frac{\beta(k') \check{y}_{kk'}}{\gamma(k')}\right) =} \nonumber \\
	& \qquad\qquad\qquad\qquad\qquad \left(1 + \check{\rho}
	 \beta(k) \right) \check{u}_k, & k \in \mathcal K  \\
	& \mathrlap{\check y_{kk'} \leq \check \eta_k; \, \check y_{kk'} \leq \check u_{k'}; \,
	\check y_{kk'} \geq \check \eta_k + \check u_{k'} - 1,} \nonumber \\
	 & &  k, k' \in \mathcal K.
    \end{align}
\end{subequations}

For ZF, we replace constraints \eqref{ZF_fair} with
\begin{subequations}
    \begin{align}
	& \check{\rho} \gamma(k) \left( \frac{1}{\check{\rho}} \sum_{k' \in \mathcal
	    K} \frac{\check{x}_{k'k}}{\gamma(k')} + \sum_{k' \in \mathcal K}
	\frac{\left( \beta(k') - \gamma(k') \right)
	\check{x}_{k'k}}{\gamma(k')}\right) = \nonumber \\
	& \qquad\qquad\qquad \left( 1 + \check{\rho}
	\left( \beta(k) - \gamma(k) \right) \right) \check{u}_k, \qquad k \in \mathcal K.
    \end{align}
\end{subequations}
Note that the order of the subscripts for $x_{kk'}$ is reversed since the
constraint index applies here to the power control coefficient $\check \eta_k$,
not the c-set variable $\check u_k$.

\bibliographystyle{IEEEtran}
\bibliography{IEEEabrv,bibliography}

\begin{thebibliography}{10}
\providecommand{\url}[1]{#1}
\csname url@samestyle\endcsname
\providecommand{\newblock}{\relax}
\providecommand{\bibinfo}[2]{#2}
\providecommand{\BIBentrySTDinterwordspacing}{\spaceskip=0pt\relax}
\providecommand{\BIBentryALTinterwordstretchfactor}{4}
\providecommand{\BIBentryALTinterwordspacing}{\spaceskip=\fontdimen2\font plus
\BIBentryALTinterwordstretchfactor\fontdimen3\font minus
  \fontdimen4\font\relax}
\providecommand{\BIBforeignlanguage}[2]{{%
\expandafter\ifx\csname l@#1\endcsname\relax
\typeout{** WARNING: IEEEtran.bst: No hyphenation pattern has been}%
\typeout{** loaded for the language `#1'. Using the pattern for}%
\typeout{** the default language instead.}%
\else
\language=\csname l@#1\endcsname
\fi
#2}}
\providecommand{\BIBdecl}{\relax}
\BIBdecl

\bibitem{gupta2015survey}
A.~Gupta and R.~K. Jha, ``A survey of {5G} network: Architecture and emerging
  technologies,'' \emph{{IEEE} access}, vol.~3, pp. 1206--1232, 2015.

\bibitem{boccardi2014five}
F.~Boccardi, R.~W. Heath, A.~Lozano, T.~L. Marzetta, and P.~Popovski, ``Five
  disruptive technology directions for {5G},'' \emph{{IEEE} Communications
  Magazine}, vol.~52, no.~2, pp. 74--80, 2014.

\bibitem{chen2014requirements}
S.~Chen and J.~Zhao, ``The requirements, challenges, and technologies for {5G}
  of terrestrial mobile telecommunication,'' \emph{{IEEE} Communications
  Magazine}, vol.~52, no.~5, pp. 36--43, 2014.

\bibitem{de2017random}
E.~De~Carvalho, E.~Bj{\"o}rnson, J.~H. S{\o}rensen, P.~Popovski, and E.~G.
  Larsson, ``Random access protocols for massive {MIMO},'' \emph{{IEEE}
  Communications Magazine}, vol.~55, no.~5, pp. 216--222, 2017.

\bibitem{bjornson2016random}
E.~Bj{\"o}rnson, E.~De~Carvalho, E.~G. Larsson, and P.~Popovski, ``Random
  access protocol for massive {MIMO}: Strongest-user collision resolution
  ({SUCR}),'' in \emph{Proc. {IEEE} International Conference on Communications
  ({ICC})}.\hskip 1em plus 0.5em minus 0.4em\relax {IEEE}, 2016, pp. 1--6.

\bibitem{sorensen2014massive}
J.~H. S{\o}rensen, E.~De~Carvalho, and P.~Popovski, ``Massive {MIMO} for crowd
  scenarios: A solution based on random access,'' in \emph{2014 Globecom
  Workshops ({GC} Wkshps)}.\hskip 1em plus 0.5em minus 0.4em\relax {IEEE},
  2014, pp. 352--357.

\bibitem{de2016random}
E.~De~Carvalho, E.~Bj{\"o}rnson, E.~G. Larsson, and P.~Popovski, ``Random
  access for massive {MIMO} systems with intra-cell pilot contamination,'' in
  \emph{2016 {IEEE} International Conference on Acoustics, Speech and Signal
  Processing ({ICASSP})}.\hskip 1em plus 0.5em minus 0.4em\relax {IEEE}, 2016,
  pp. 3361--3365.

\bibitem{marzetta2010noncooperative}
T.~L. Marzetta, ``Noncooperative cellular wireless with unlimited numbers of
  base station antennas,'' \emph{{IEEE} Transactions on Wireless
  Communications}, vol.~9, no.~11, pp. 3590--3600, 2010.

\bibitem{larsson2014massive}
E.~G. Larsson, O.~Edfors, F.~Tufvesson, and T.~L. Marzetta, ``Massive {MIMO}
  for next generation wireless systems,'' \emph{{IEEE} communications
  magazine}, vol.~52, no.~2, pp. 186--195, 2014.

\bibitem{harris2016serving}
P.~Harris, W.~B. Hasan, S.~Malkowsky, J.~Vieira, S.~Zhang, M.~Beach, L.~Liu,
  E.~Mellios, A.~Nix, S.~Armour \emph{et~al.}, ``Serving 22 users in real-time
  with a 128-antenna massive {MIMO} testbed,'' in \emph{Proc. 2016 IEEE
  International Workshop on Signal Processing Systems ({SiPS})}.\hskip 1em plus
  0.5em minus 0.4em\relax {IEEE}, 2016, pp. 266--272.

\bibitem{zheng2014massive}
K.~Zheng, S.~Ou, and X.~Yin, ``Massive {MIMO} channel models: A survey,''
  \emph{International Journal of Antennas and Propagation}, vol. 2014, 2014.

\bibitem{araujo2016massive}
D.~C. Ara{\'u}jo, T.~Maksymyuk, A.~L. de~Almeida, T.~Maciel, J.~C. Mota, and
  M.~Jo, ``Massive {MIMO}: survey and future research topics,'' \emph{{IET}
  Communications}, vol.~10, no.~15, pp. 1938--1946, 2016.

\bibitem{payami2012channel}
S.~Payami and F.~Tufvesson, ``Channel measurements and analysis for very large
  array systems at 2.6 {GHz},'' in \emph{Proc. 2012 6th European Conference on
  Antennas and Propagation ({EUCAP})}.\hskip 1em plus 0.5em minus 0.4em\relax
  {IEEE}, 2012, pp. 433--437.

\bibitem{gao2012measured}
X.~Gao, F.~Tufvesson, O.~Edfors, and F.~Rusek, ``Measured propagation
  characteristics for very-large {MIMO} at 2.6 {GHz},'' in \emph{2012
  Conference Record of the Forty Sixth Asilomar Conference on Signals, Systems
  and Computers ({ASILOMAR})}.\hskip 1em plus 0.5em minus 0.4em\relax {IEEE},
  2012, pp. 295--299.

\bibitem{hoydis2013massive}
J.~Hoydis, S.~Ten~Brink, and M.~Debbah, ``Massive {MIMO} in the {UL/DL} of
  cellular networks: How many antennas do we need?'' \emph{{IEEE} Journal on
  selected Areas in Communications}, vol.~31, no.~2, pp. 160--171, 2013.

\bibitem{yin2013coordinated}
H.~Yin, D.~Gesbert, M.~Filippou, and Y.~Liu, ``A coordinated approach to
  channel estimation in large-scale multiple-antenna systems,'' \emph{{IEEE}
  Journal on selected areas in communications}, vol.~31, no.~2, pp. 264--273,
  2013.

\bibitem{masouros2013large}
C.~Masouros, M.~Sellathurai, and T.~Ratnarajah, ``Large-scale {MIMO}
  transmitters in fixed physical spaces: The effect of transmit correlation and
  mutual coupling,'' \emph{{IEEE} Transactions on Communications}, vol.~61,
  no.~7, pp. 2794--2804, 2013.

\bibitem{rusek2013scaling}
F.~Rusek, D.~Persson, B.~K. Lau, E.~G. Larsson, T.~L. Marzetta, O.~Edfors, and
  F.~Tufvesson, ``Scaling up {MIMO}: Opportunities and challenges with very
  large arrays,'' \emph{{IEEE} signal processing magazine}, vol.~30, no.~1, pp.
  40--60, 2013.

\bibitem{hochwald2004multiple}
B.~M. Hochwald, T.~L. Marzetta, and V.~Tarokh, ``Multiple-antenna channel
  hardening and its implications for rate feedback and scheduling,''
  \emph{{IEEE} Transactions on Information Theory}, vol.~50, no.~9, pp.
  1893--1909, 2004.

\bibitem{hoydis2011massive}
J.~Hoydis, S.~Ten~Brink, and M.~Debbah, ``Massive {MIMO}: How many antennas do
  we need?'' in \emph{Proc. 2011 49th Annual Allerton Conference on
  Communication, Control, and Computing}.\hskip 1em plus 0.5em minus
  0.4em\relax {IEEE}, 2011, pp. 545--550.

\bibitem{ngo2013energy}
H.~Q. Ngo, E.~G. Larsson, and T.~L. Marzetta, ``Energy and spectral efficiency
  of very large multiuser {MIMO} systems,'' \emph{{IEEE} Transactions on
  Communications}, vol.~61, no.~4, pp. 1436--1449, 2013.

\bibitem{yang2013performance}
H.~Yang and T.~L. Marzetta, ``Performance of conjugate and zero-forcing
  beamforming in large-scale antenna systems,'' \emph{{IEEE} Journal on
  Selected Areas in Communications}, vol.~31, no.~2, pp. 172--179, 2013.

\bibitem{bjornson2016massiveb}
E.~Bj{\"o}rnson, E.~G. Larsson, and M.~Debbah, ``Massive {MIMO} for maximal
  spectral efficiency: How many users and pilots should be allocated?''
  \emph{{IEEE} Transactions on Wireless Communications}, vol.~15, no.~2, pp.
  1293--1308, 2016.

\bibitem{bjornson2016massivea}
E.~Bj{\"o}rnson, E.~G. Larsson, and T.~L. Marzetta, ``Massive {MIMO}: Ten myths
  and one critical question,'' \emph{{IEEE} Communications Magazine}, vol.~54,
  no.~2, pp. 114--123, 2016.

\bibitem{saxena2015mitigating}
V.~Saxena, G.~Fodor, and E.~Karipidis, ``Mitigating pilot contamination by
  pilot reuse and power control schemes for massive {MIMO} systems,'' in
  \emph{Proc. 2015 {IEEE} 81st Vehicular Technology Conference (VTC
  Spring)}.\hskip 1em plus 0.5em minus 0.4em\relax {IEEE}, 2015, pp. 1--6.

\bibitem{zappone2016energy}
A.~Zappone, L.~Sanguinetti, G.~Bacci, E.~Jorswieck, and M.~Debbah,
  ``Energy-efficient power control: A look at {5G} wireless technologies,''
  \emph{{IEEE} Transactions on Signal Processing}, vol.~64, no.~7, pp.
  1668--1683, 2016.

\bibitem{choi2014massive}
J.~Choi, ``Massive {MIMO} with joint power control,'' \emph{{IEEE} Wireless
  Communications Letters}, vol.~3, no.~4, pp. 329--332, 2014.

\bibitem{bogale2014pilot}
T.~E. Bogale and L.~B. Le, ``Pilot optimization and channel estimation for
  multiuser massive {MIMO} systems,'' in \emph{Proc. 2014 48th Annual
  Conference on Information Sciences and Systems ({CISS})}.\hskip 1em plus
  0.5em minus 0.4em\relax {IEEE}, 2014, pp. 1--6.

\bibitem{bjornson2015optimal}
E.~Bj{\"o}rnson, L.~Sanguinetti, J.~Hoydis, and M.~Debbah, ``Optimal design of
  energy-efficient multi-user {MIMO} systems: Is massive {MIMO} the answer?''
  \emph{{IEEE} Transactions on Wireless Communications}, vol.~14, no.~6, pp.
  3059--3075, 2015.

\bibitem{van2016joint}
T.~Van~Chien, E.~Bj{\"o}rnson, and E.~G. Larsson, ``Joint power allocation and
  user association optimization for massive {MIMO} systems,'' \emph{{IEEE}
  Transactions on Wireless Communications}, vol.~15, no.~9, pp. 6384--6399,
  2016.

\bibitem{huh2011multi}
H.~Huh, S.-H. Moon, Y.-T. Kim, I.~Lee, and G.~Caire, ``Multi-cell {MIMO}
  downlink with cell cooperation and fair scheduling: A large-system limit
  analysis,'' \emph{{IEEE} Transactions on Information Theory}, vol.~57,
  no.~12, pp. 7771--7786, 2011.

\bibitem{guozhen2014joint}
X.~Guozhen, L.~An, J.~Wei, X.~Haige, and L.~Wu, ``Joint user scheduling and
  antenna selection in distributed massive {MIMO} systems with limited backhaul
  capacity,'' \emph{China Communications}, vol.~11, no.~5, pp. 17--30, 2014.

\bibitem{benmimoune2015joint}
M.~Benmimoune, E.~Driouch, W.~Ajib, and D.~Massicotte, ``Joint transmit antenna
  selection and user scheduling for massive {MIMO} systems,'' in \emph{Proc.
  2015 IEEE Wireless Communications and Networking Conference ({WCNC})}.\hskip
  1em plus 0.5em minus 0.4em\relax {IEEE}, 2015, pp. 381--386.

\bibitem{xu2014user}
Y.~Xu, G.~Yue, and S.~Mao, ``User grouping for massive {MIMO} in {FDD} systems:
  New design methods and analysis,'' \emph{{IEEE} Access}, vol.~2, pp.
  947--959, 2014.

\bibitem{Bjorklund2003}
P.~Bj{\"o}rklund, P.~V{\"a}rbrand, and D.~Yuan, ``{Resource Optimization of
  Spatial {TDMA} in Ad hoc Radio Networks: a Column Generation Approach},'' in
  \emph{Proc. {IEEE INFOCOM}}, March 2003, pp. 818--824.

\bibitem{CaC2006}
A.~Capone and G.~Carello, ``{Scheduling Optimization in Wireless MESH Networks
  with Power Control and Rate Adaptation},'' in \emph{Proc. IEEE SECON 2006},
  Sept. 25-28 2006.

\bibitem{Pioro2011}
M.~Pi\'{o}ro, M.~\.{Z}otkiewicz, B.~Staehle, D.~Staehle, and D.~Yuan, ``{On
  Max-min Fair Flow Optimization in Wireless Mesh Networks},'' \emph{Ad Hoc
  Networks}, vol.~13, no.~0, pp. 134--152, 2014.

\bibitem{Yuan-CS-heu-2015}
Y.~Li, M.~Pi\'{o}ro, D.~Yuan, and J.~Su, ``{Optimizing Link Rate Assignment and
  Transmission Scheduling in WMN through Compatible Set Generation},''
  \emph{Telecommunications Systems Journal}, no.~61, pp. 325--335, 2016.

\bibitem{Angelakis2014}
V.~Angelakis, A.~Ephremides, Q.~He, and D.~Yuan, ``{Minimum-Time Link
  Scheduling for Emptying Wireless Systems: Solution Characterization and
  Algorithmic Framework},'' \emph{IEEE Transactions on Information Theory},
  vol.~60, no.~2, pp. 1083--1100, 2014.

\bibitem{Capone2010}
A.~Capone, G.~Carello, I.~Filippini, S.~Gualandi, and F.~Malucelli, ``{Routing,
  Scheduling and Channel Assignment in Wireless Mesh Networks: Optimization
  Models and Algorithms},'' \emph{Ad Hoc Networks}, vol.~8, no.~6, pp.
  545--563, 2010.

\bibitem{PTC-2018}
M.~Pi\'{o}ro, A.~Tomaszewski, and A.~Capone, ``Maximization of multicast
  periodic traffic throughput in multi-hop wireless networks with broadcast
  transmissions,'' \emph{Ad Hoc Networks}, vol.~77, pp. 119--142, 2018.

\bibitem{Zotkiewicz2014}
M.~\.Zotkiewicz, ``{Optimal rate selection and scheduling in WMN with
  interference cancelation using overheard transmissions},'' \emph{Journal of
  Applied Mathematics}, 2014.

\bibitem{Capone2014}
Y.~Li, A.~Capone, and D.~Yuan, ``{On End-to-end Delay Minimization in Wireless
  Network under Physical Interference Model},'' in \emph{Proc. IEEE INFOCOM},
  2015, pp. 2020--2028.

\bibitem{Capone2015-inoc}
A.~Capone, M.~Pi\'{o}ro, Y.~Li, and D.~Yuan, ``{On Packet Transmission
  Scheduling for Min-Max Delay and Energy Consumption in Wireless Mesh Sensor
  Networks},'' in \emph{Proc. International Network Optimization Conference},
  May 2015.

\bibitem{fitzgerald2018energy}
E.~Fitzgerald, M.~Pi{\'o}ro, and A.~Tomaszewski, ``Energy-optimal data
  aggregation and dissemination for the {Internet of Things},'' \emph{{IEEE}
  {Internet of Things} Journal}, vol.~5, no.~2, pp. 955--969, 2018.

\bibitem{marzetta2016fundamentals}
T.~L. Marzetta, E.~G. Larsson, H.~Yang, and H.~Q. Ngo, \emph{Fundamentals of
  Massive MIMO}.\hskip 1em plus 0.5em minus 0.4em\relax Cambridge University
  Press, 2016.

\bibitem{foukas2017network}
X.~Foukas, G.~Patounas, A.~Elmokashfi, and M.~K. Marina, ``Network slicing in
  {5G}: Survey and challenges,'' \emph{{IEEE} Communications Magazine},
  vol.~55, no.~5, pp. 94--100, 2017.

\bibitem{korte-vygen12}
B.~Korte and J.~Vygen, \emph{{Combinatorial Optimization -- Theory and
  Algorithms}}.\hskip 1em plus 0.5em minus 0.4em\relax Springer, 2012.

\bibitem{ahuja1993}
R.~Ahuja, T.~Magnanti, and J.~Orlin, \emph{{Network Flows: Theory, Algorithms,
  and Applications}}.\hskip 1em plus 0.5em minus 0.4em\relax Prentice Hall,
  1993.

\bibitem{lasdon1970}
L.~Lasdon, \emph{{Optmization Theory for Large Systems}}.\hskip 1em plus 0.5em
  minus 0.4em\relax Macmillan, 1970.

\bibitem{minoux1986}
M.~Minoux, \emph{{Mathematical Programming -- Theory and Algorithms}}.\hskip
  1em plus 0.5em minus 0.4em\relax J. Wiley \& Sons, 1986.

\bibitem{pioro2012network}
M.~Pi\'{o}ro, ``Network optimization techniques,'' in \emph{Mathematical
  Foundations for Signal Processing, Communications, and Networking},
  E.~Serpedin, T.~Chen, and D.~Rajan, Eds.\hskip 1em plus 0.5em minus
  0.4em\relax Boca Raton, USA: CRC Press, 2012, ch.~18, pp. 627--690.

\bibitem{Goussevskaia2007}
O.~Goussevskaia, Y.~A. Oswald, and R.~Wattenhofer, ``{Complexity in Geometric
  SINR},'' in \emph{ACM Mobihoc}, 2007, pp. 100--109.

\bibitem{nemhauser-wolsey}
G.~Nemhauser and L.~Wolsey, \emph{{Integer and Combinatorial
  Optimization}}.\hskip 1em plus 0.5em minus 0.4em\relax J. Wiley \& Sons,
  1999.

\bibitem{nptz2013}
D.~Nace, M.~Pi\'{o}ro, A.~Tomaszewski, and M.~\.{Z}otkiewicz, ``{Complexity of
  a Classical Flow Restoration Problem},'' \emph{Networks: an International
  Journal}, vol.~62, no.~2, pp. 149--160, 2013.

\bibitem{malkowsky2017world}
S.~Malkowsky, J.~Vieira, L.~Liu, P.~Harris, K.~Nieman, N.~Kundargi, I.~C. Wong,
  F.~Tufvesson, V.~{\"O}wall, and O.~Edfors, ``The world’s first real-time
  testbed for massive {MIMO}: Design, implementation, and validation,''
  \emph{IEEE Access}, vol.~5, pp. 9073--9088, 2017.

\bibitem{gunnarsson2018channel}
S.~Gunnarsson, J.~Flordelis, L.~V. der Perre, and F.~Tufvesson, ``Channel
  hardening in massive {MIMO} --- a measurement based analysis,'' in
  \emph{Proc. {IEEE} International Workshop on Signal Processing Advances in
  Wireless Communications ({SPAWC})}, June 2018, pp. 1--5.

\bibitem{report}
E.~Fitzgerald, M.~Pi\'{o}ro, and F.~Tufvesson, ``Massive {MIMO} optimization
  with compatible sets,'' Department of Electrical and Information Technology,
  Lund University, Sweden, Tech. Rep., 2019, {ISBN}: 978-91-7895-079-9.

\end{thebibliography}

\end{document}